\documentclass[authoryear,preprint]{elsarticle}
\usepackage{natbib}
\usepackage{lineno, hyperref}

\usepackage{amssymb,amsmath}
\usepackage{bm,upgreek}
\usepackage{graphics}
\usepackage{amsfonts}
\usepackage{color}
\usepackage{amsthm}

\usepackage[symbol]{footmisc} 

\theoremstyle{remark}

\graphicspath{%
{Fig/}%
{Fig_SA/}%
}

\def\inclps#1#2#3{\resizebox{#1}{#2}{\includegraphics{#3}}}

\newcommand\rd{{\rm{d}}}
\newcommand{\bfm}[1]{{\rm\bf #1}}

\DeclareMathOperator{\sign}{sign}

\definecolor{myviolet}{RGB}{255,0,255}
\definecolor{darkgreen}{rgb}{0,0.5,0}

\let\oldequation\equation
\let\oldendequation\endequation
\renewenvironment{equation}
  {\linenomathNonumbers\oldequation}
  {\oldendequation\endlinenomath}

\begin{document}

\begin{frontmatter}
\title{Twin branching in shape memory alloys: a 1D model with energy dissipation effects$^\dagger$}

\author[IPPT]{Stanis{\l}aw Stupkiewicz\corref{cor1}}
\ead{sstupkie@ippt.pan.pl}
\author[IPPT]{Seyedshoja Amini}
\ead{samini@ippt.pan.pl}
\author[IPPT]{Mohsen Rezaee-Hajidehi}
\ead{mrezaee@ippt.pan.pl}

\cortext[cor1]{Corresponding author}

\address[IPPT]{Institute of Fundamental Technological Research, Polish Academy of Sciences,\\
Pawi\'nskiego 5B, 02-106 Warsaw, Poland.}

\begin{abstract}
We develop a 1D model of twin branching in shape memory alloys. The free energy of the branched microstructure comprises the interfacial and elastic strain energy contributions, both expressed in terms of the average twin spacing treated as a continuous function of the position. The total free energy is then minimized, and the corresponding Euler--Lagrange equation is solved numerically using the finite element method. The model can be considered as a continuous counterpart of the recent discrete model of \citet{seiner2020branching}, and our results show a very good agreement with that model in the entire range of physically relevant parameters. Furthermore, our continuous setting facilitates incorporation of energy dissipation into the model. The effect of rate-independent dissipation on the evolution of the branched microstructure is thus studied. The results show that significant effects on the microstructure and energy of the system are expected only for relatively small domain sizes.
\footnotetext[2]{Published in \emph{Eur. J. Mech. A/Solids}, 
2025, doi: 10.1016/j.euromechsol.2025.105671}
\end{abstract}


\end{frontmatter}


\section{Introduction}

Martensitic transformation, which is the basic deformation mechanism in shape memory alloys, proceeds through formation and evolution of microstructure \citep{Bhattacharya2003}. 
Interfaces between the product phase (martensite) and the parent phase (austenite) constitute the most common element of the microstructure. Kinematic compatibility of the two phases at such interfaces is typically achieved through twinning of martensite so that martensite appears in the form of a laminate of two twin-related variants. 
Twin branching, which is the subject of this work, is a phenomenon frequently observed in experiments \citep[e.g.,][]{chu1993hysteresis,bronstein2019analysis,qin2023compatibility,zhang2024austenite} in which the thickness of the individual twins is gradually reduced towards the austenite--martensite interface. 

The phenomenon of twin branching can be explained by the propensity of the material to minimize its total free energy which includes the interfacial energy of twin boundaries and the energy of elastic strains that accommodate the local incompatibility of the phases at the austenite--martensite interface. The latter energy is reduced when the twin spacing is decreased (close to the austenite--martensite interface), while the former energy is reduced when the twin spacing is increased (far from the austenite--martensite interface). Accordingly, the total energy can be minimized by a microstructure with the twin spacing varying as a function of the distance from the austenite--martensite interface, and this is achieved through twin branching. Note that branching itself incurs an energetic cost that participates in the competition between the free energy contributions. 

Twin branching is related, through the energy-minimization background and sometimes also through the appearance, to other phenomena observed in natural and engineered materials. These include branching in magnetic domains \citep{hubert1998magnetic}, magnetic flux structures in superconductors \citep{huebener2013magnetic}, hierarchical wrinkling patterns in thin films \citep{vandeparre2010hierarchical,ma2020tunable}, and others.

The theoretical analysis of twin branching started with the seminal contributions of \citet{kohn1992branching,kohn1994surface}, followed by many related works, mostly in the mathematics community \citep[e.g.,][]{conti2000branched,capella2009rigidity,chan2015energy,dondl2016optimization,simon2021rigidity}.
This line of research delivers, in particular, the scaling laws for energy and for characteristic dimensions of the branched microstructure. 
As a follow up, a quantitative model of twin branching in shape memory alloys has been recently developed by \citet{seiner2020branching}. 
The model assumes that the branched microstructure consists of a family of self-similar branching generations, as in several earlier models, and employs refined estimates of the elastic strain energy of branching that have been derived for the considered microstructure. The parameters characterizing the actual microstructure are then obtained by minimizing the total free energy. This model will be used as a reference for our model developed in this work. 

Direct spatially-resolved simulation of a branched microstructure by using, for example, the phase-field method, is not feasible because of the multiscale nature of the branched microstructure. 
However, some features characteristic for twin branching can be observed in phase-field simulations \citep[e.g.,][]{finel2010phase,tuuma2016phase,amini2023energy}. 
These are mostly limited to tip splitting of martensite needles, or the like, and thus correspond to just a single branching generation. To the best of our knowledge, spatially-resolved simulations of hierarchical branched microstructures have not been reported to date. 
A kind of hierarchical twin branching is reported in the related work of \citet{li1999theory}, but the interfacial energy is not included in that model (only the elastic strain energy is minimized) so that the model does not capture the essential competition between the interfacial and elastic strain energy contributions. 

In this paper, we propose a new approach to the modeling of twin branching. A 1D model is developed in which the twin spacing is treated as a continuous function of the position.
In this setting, the free energy of the system is formulated which includes three contributions: the interfacial energy of twin boundaries, the elastic strain energy of branching, and the energy of elastic strains that accommodate the local incompatibility at the austenite--twinned martensite interface. 
The total free energy is then minimized, thus delivering the governing equation (ODE) for the twin spacing that is solved numerically using the finite element method. 
The model can be considered a continuous counterpart of the discrete model of \citet{seiner2020branching}, and we show that the predictions of the two models agree very well for a wide range of model parameters. 

We call our model a continuous model since the twin spacing is treated as a continuous function of the position and the governing equation can be formulated as an ODE. This is in contrast to the model of \citet{seiner2020branching}, which we call a discrete model, since that model considers a discrete number of branching generations thus leading to a step-like dependence of the twin spacing on the position. Moreover, the solution in that model is obtained from a discrete minimization problem, in particular, involving minimization with respect to the (integer) number of branching generations. 

A model similar to our model has been recently presented by \citet{zhang2024austenite} for the analysis of experimental data on austenite--martensite interfacial patterns in NiMnGa reported in the reference. 
That model also considers the twin spacing as a continuous function of the position and adopts the same form of the free energy density in the branched martensite as in our model. The difference is that our model includes the energy of elastic micro-strains at the A--MM interface, which is disregarded in the model of \citet{zhang2024austenite}. This leads to different boundary conditions for the corresponding ODE. The model of \citet{zhang2024austenite}, once fitted to their experimental data, correctly describes the profile of the twin spacing in the vicinity of the austenite--martensite interface.

Secondly, for the first time, we consider the energy dissipation associated with the evolution of the branched microstructure. 
This is largely facilitated by our continuous formulation (note that development of the respective discrete formulation of the evolution problem may not be readily feasible). 
The evolution equation for the twin spacing is obtained by minimization of the rate potential that comprises the rate of the total free energy and the rate-independent dissipation potential. 
Finite-element computations are then carried out for a high but realistic value of the rate-independent threshold for propagation of twin interfaces. The results suggest that the energy dissipation effects are small or even negligible in most cases. Only for a relatively high value of the dimensionless interfacial energy (which corresponds, for instance, to a small size of the twinned domain) and for a high interface propagation threshold, a significant effect on both the microstructure (twin spacing) and energy can be observed.

The remainder of the paper is organized as follows. In Section~\ref{sec:background}, the discrete model of twin branching of \citet{seiner2020branching} is briefly described as a reference for our subsequent developments. In Section~\ref{sec:model}, the continuous model of twin branching is developed and sample results of finite-element computations are presented, including a comparison with the predictions of the model of \citet{seiner2020branching}. As a side note, an analytical solution to this (static) problem is discussed in \ref{app:analytical}. In Section~\ref{sec:dissip}, energy dissipation is introduced into the model and the corresponding effects are studied for a few representative cases of evolution of the branched microstructure. 
Technical details concerning the derivation of the evolution problem and its computational treatment are presented in \ref{app:speed} and \ref{app:AL}, respectively.
A~general discussion on the model and its applicability is presented in Section~\ref{sec:discussion}, followed by concluding remarks in Section~\ref{sec:conclusion}.

\section{Background}
\label{sec:background}

\subsection{Austenite--twinned martensite interface}
\label{sec:CTM}

In majority of shape memory alloys, the interface between the austenite and the martensite forms such that the martensite is internally twinned, i.e., it has the form of a fine laminate of two twin-related variants of martensite. This is because a single martensite variant is usually not compatible with the austenite and thus cannot form a low-energy (stress-free) austenite--martensite interface \citep{Bhattacharya2003}. 
The orientation and other features of the austenite--twinned martensite (A--MM) interface can be predicted by using the crystallographic theory of martensite \citep{wechsler1953onthetheory,ball1987fine,Bhattacharya2003}, which relies on the kinematic compatibility conditions formulated for the twin interfaces and for the A--MM interface.

Consider thus two variants of martensite characterized by the transformation stretch tensors $\bfm{U}_A$ and $\bfm{U}_B$. In stress-free conditions, the condition of kinematic compatibility of the two variants along a planar interface of the unit normal $\bfm{n}$ can be written in the form of the following \emph{twinning equation},
	\begin{equation}
		\label{eq:twinning}
		\bfm{R} \bfm{U}_A - \bfm{U}_B = \hat{\bfm{a}} \otimes \bfm{n} ,
	\end{equation}
where $\hat{\bfm{a}}$ is a non-zero vector and $\bfm{R}$ is a rotation tensor. 
Consider further a simple laminate formed by the two martensite variants with $\lambda$ denoting the volume fraction of the variant $\bfm{U}_A$. The \emph{overall} compatibility of the austenite and twin laminate is then expressed in the form of the \emph{habit plane equation},
	\begin{equation}
		\label{eq:habit}
		\bfm{Q} ( \lambda \bfm{R} \bfm{U}_A + (1-\lambda) \bfm{U}_B ) - \bfm{I} = \bfm{b} \otimes \bfm{m} ,
	\end{equation}
where $\bfm{m}$ is the unit normal to the (planar) A--MM interface, $\bfm{b}$ is a non-zero vector, $\bfm{Q}$ is a rotation tensor, and $\bfm{I}$ is the unit tensor that corresponds to the deformation gradient of the unstressed austenite.
The unknown $\bfm{R}$, $\hat{\bfm{a}}$, $\bfm{n}$, $\bfm{Q}$, $\bfm{b}$, $\bfm{m}$, and $\lambda$ can be found using the procedure described, for instance, in \citet{Bhattacharya2003}. 

It is stressed that the habit plane equation~\eqref{eq:habit} describes the overall compatibility between the twin laminate and the austenite, while the individual variants of martensite are not compatible with the austenite. Accordingly, elastic strains are needed to accommodate the corresponding local incompatibility, hence a transition layer with non-zero elastic strains necessarily forms along the A--MM interface. 
The energy of the transition layer depends on the characteristic spacing of the twin laminate (twin spacing) and is, to the first order, proportional to it. This energy can thus be lowered by reducing the twin spacing. However, a finely twinned laminate would have a high density of twin interfaces and thus a high total energy of twin boundaries, since each twin interface has a non-zero energy proportional to the area of the interface. The actual twin spacing minimizing the total energy would thus result from the competition between the elastic micro-strain energy in the transition layer and the total interfacial energy of twin boundaries \citep{khachaturyan1983theory}. 
The total energy can be further minimized through formation of a branched microstructure such that twins are (relatively) coarse far from the A--MM interface to reduce the interfacial energy and are fine close to the A--MM interface to reduce the elastic micro-strain energy. The branching itself is associated with the energy of elastic strains needed to accommodate the incompatibility induced by the deviation from the fully compatible twin interfaces, and this energy also contributes to the total energy balance.

\subsection{Discrete model of twin branching}

\begin{figure}
	\centerline{\inclps{1.1\textwidth}{!}{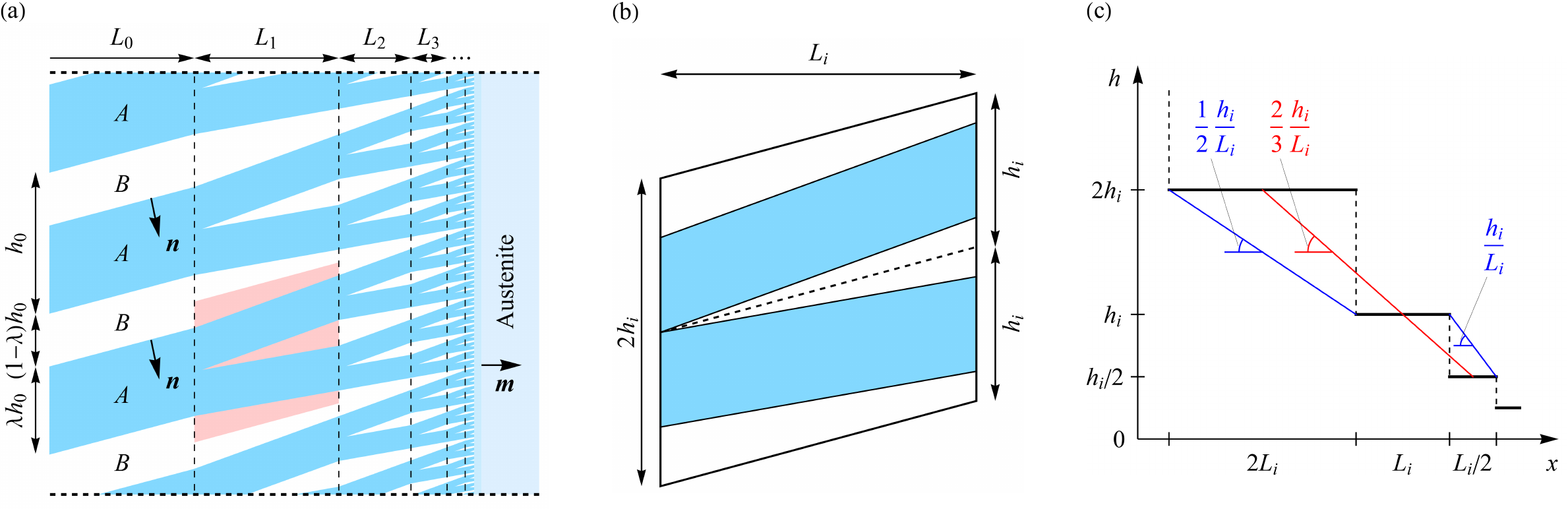}}
	\caption{Discrete model of twin branching \citep{seiner2020branching}: (a) self-similar microstructure with $N$ branching generations; 
		(b) unit cell; (c) twin spacing $h$ as a function of the position $x$. 
        Possible ways to determine the derivative of a continuous function $h(x)$ are shown in panel (c).}
	\label{fig:Seiner}
\end{figure}

A discrete model of twin branching has been recently developed by \citet{seiner2020branching}, and this model is briefly summarized below. The model is based on the minimization of the total free energy of the specific branching microstructure shown in Fig.~\ref{fig:Seiner}(a). It is assumed that the twin spacing is refined through a sequence of self-similar branching segments. 
Each branching generation, indexed by $i$, is formed by a layer of the corresponding unit cells (segments) that are repeated in a periodic manner with $L_i$ denoting the length measured in the direction perpendicular to the A--MM interface and $h_i$ denoting the twin spacing measured along the A--MM interface, while the width of the segment is equal to $2h_i$, see Fig.~\ref{fig:Seiner}(b). Here and in the following, we keep the notation used by \citet{seiner2020branching}, except that the twin spacing is denoted by $h_i$ rather than by $d_i$ and $\Gamma$ replaces $G_{AB}$. 
Note that the first layer of the length $L_0$ and spacing $h_0$ is not branched, see Fig.~\ref{fig:Seiner}(a).

For a given total length $L$ of the twinned domain, the microstructure of the branched twin laminate is characterized by the number of branching generations $N$, the twin spacing $h_0$ in the first layer, and the individual lengths $L_i$. These variables are determined by minimizing the total free energy $E$,
	\begin{equation}
		\label{eq:minE}
		E = \frac1{h_0} \left( E_{\rm surf}^{(0)} +
            \sum_{i=1}^N 2^{i-1} \big( E_{\rm elast}^{(i)} + E_{\rm surf}^{(i)} \big) \right) + \Gamma h_N
		\;\;\; \to \;\;\; \min ,
	\end{equation}
where $E_{\rm elast}^{(i)}$ denotes the elastic strain energy of one segment in the $i$-th branching layer, $E_{\rm surf}^{(i)}$ denotes the interfacial energy of the twin boundaries contained within that segment, and the lengths $L_i$ must satisfy the constraint $L=\sum_{i=0}^N L_i$. 
The last term represents the elastic micro-strain energy in the transition layer along the A--MM interface with $\Gamma$ denoting the energy factor and $h_N=h_0/2^N$ denoting the twin spacing at the A--MM interface. Considering that the first layer ($i=0$) is not branched, the corresponding elastic strain energy is equal to zero, $E_{\rm elast}^{(0)}=0$.

For the microstructure of the branching segment shown in Fig.~\ref{fig:Seiner}(b), an upper-bound estimate of the elastic strain energy of branching has been developed in the following form, see Eq.~(S.6) in the supplementary material of \citet{seiner2020branching},
	\begin{equation}
		\label{eq:Eel}
		E_{\rm elast}^{(i)} = \frac14 \mu \lambda (1-\lambda) g_2 L_i h_i \varepsilon_i^2 , \qquad
		\varepsilon_i = \frac{(1-\lambda)\alpha h_i}{L_i} ,
	\end{equation}
where
	\begin{equation}
		\label{eq:alpha}
		\alpha = \sqrt{1-(\bfm{m}\cdot\bfm{n})^2} , \qquad
		g_2 = \frac12 (|\bfm{a}|^2+(\bfm{a}\cdot\bfm{m})^2) ,
	\end{equation}
and $\mu$ denotes the shear modulus. Here, $\lambda$, $\bfm{m}$, $\bfm{n}$, and $\bfm{a}=\bfm{Q}\hat{\bfm{a}}$ follow from the crystallographic theory outlined in Section~\ref{sec:CTM}. 
The above estimate corresponds to the case of elastic isotropy. A more refined estimate considering elastic anisotropy is given by Eq.~(30) in \citet{seiner2020branching}.
The energy $E_{\rm elast}^{(i)}$ refers to a unit length in the out-of-plane direction, and the density per unit volume is thus obtained by dividing $E_{\rm elast}^{(i)}$ by the area of the segment,
	\begin{equation}
		\label{eq:elastic}
		\frac{E_{\rm elast}^{(i)}}{2 h_i L_i} = \frac18 \mu \lambda (1-\lambda)^3 \alpha^2 g_2 \left( \frac{h_i}{L_i} \right)^2 .
	\end{equation}
This result will be used to calibrate our continuous model in Section~\ref{sec:model}.

{The interfacial energy $E_{\rm surf}^{(i)}$ of branched segments ($i\geq1$) has been estimated as, see Eq.~(S.9) in the supplementary material of \citet{seiner2020branching},
    \begin{equation}
        \label{eq:Eint:full}
        E_{\rm surf}^{(i)} = 2 \gamma (1-\lambda) h_i \left( \frac{2}{\varepsilon_i} + \lambda + \frac{1+\lambda}{2} \varepsilon_i \right) ,
    \end{equation}
where $\gamma$ is a material parameter characterizing the interfacial energy of twin boundaries (density per unit area). In this model, the actual energy of each interface is assumed to depend on the jump of the deformation gradient at this interface, which accounts, in a simple manner, for the dependence of the interfacial energy on the interface orientation. The above expression accounts also for the changes in the area (length) of the twin interfaces within each segment as a function of the aspect ratio $2h_i/L_i$ of the segment. As discussed in Section~\ref{sec:discussion}, the effects related to the changes in the local interfacial energy are not significant in the considered case of a CuAlNi alloy and can be disregarded.}

Assuming {thus} that $L_i\gg h_i$, the total interfacial energy $E_{\rm surf}^{(i)}$ {of four interfaces of length $L_i/\alpha$ and the corresponding} density can be simply estimated as
	\begin{equation}
		\label{eq:Eint}
		E_{\rm surf}^{(i)} = \frac{4 L_i \gamma}{\alpha} , \qquad
		\frac{E_{\rm surf}^{(i)}}{2 h_i L_i} = \frac{2 \gamma}{\alpha h_i} .
	\end{equation}
Note that the nominal twinning planes (of normal $\bfm{n}$) are not necessarily perpendicular to the A--MM interface (of normal $\bfm{m}$), see Fig.~\ref{fig:Seiner}, hence the factor $\alpha$ in Eq.~\eqref{eq:Eint}, see also Eq.~\eqref{eq:alpha}${}_1$. 
In fact, Eq.~\eqref{eq:Eint}${}_1$ can be obtained from Eq.~\eqref{eq:Eint:full} by substituting $\varepsilon_i$ from Eq.~\eqref{eq:Eel}${}_2$ and by setting $h_i/L_i=0$, which amounts to disregarding the second and third term in the parentheses in Eq.~\eqref{eq:Eint:full}. 
Since there are two twin interfaces in each segment of the first layer ($i=0$), we have $E_{\rm surf}^{(0)} = 2 L_0 \gamma/\alpha$. 
{In} our model, we will only consider the simple estimate in Eq.~\eqref{eq:Eint}.

The elastic micro-strain energy factor $\Gamma$ in Eq.~\eqref{eq:minE}, for which \citet{seiner2020branching} developed a separate construction and derived the corresponding upper-bound estimate of the elastic strain energy, is discussed later.

\section{Continuous model of twin branching}
\label{sec:model}

\subsection{Twin spacing as a continuous function}

In our model, the twin spacing $h$ is considered to be a continuous function of the position $x$, thus $h=h(x)$ for $0\leq x\leq L$.
Considering that real microstructures do not exhibit periodicity nor perfect arrangement of branching generations, as assumed in the discrete model of \citet{seiner2020branching}, the twin spacing $h(x)$ should be interpreted as the average spacing measured over a representative part of the A--MM interface and determined at a given distance from it. 
Note that, in accord with \citet{seiner2020branching}, the twin spacing $h(x)$ is here measured in the direction parallel to the A--MM interface. The spatial coordinate $x$ is measured in the direction perpendicular to the A--MM interface with $x=0$ at the free boundary and $x=L$ at the A--MM interface, see Fig.~\ref{fig:Seiner}(a).
{As in the discrete model, the free boundary is assumed parallel to the A--MM interface, hence the length $L$ is independent of the position along the A--MM interface. This assumption can be relaxed, as discussed in Section~\ref{sec:discussion}.}

The estimate of the elastic strain energy of branching, Eq.~\eqref{eq:Eel}, will be used to characterize the elastic strain energy in the continuous setting. To this end, we note that the energy density within a single cell depends on the ratio $h_i/L_i$, see Eq.~\eqref{eq:elastic}. 
Now, referring to Fig.~\ref{fig:Seiner}(c), the spatial derivative of $h(x)$ can be related to the ratio $h_i/L_i$ through
	\begin{equation}
		\label{eq:dhdx}
		h_{,x} = \frac{\rd h}{\rd x} \approx \frac{\Delta h}{\Delta x} \approx \frac{\frac12 h_i-2h_i}{\frac14 L_i+2 L_i} = -\frac{2h_i}{3L_i} .
	\end{equation}
By combining Eqs.~\eqref{eq:elastic} and~\eqref{eq:dhdx}, the elastic strain energy density can thus be expressed as a function of the (squared) spatial derivative $h_{,x}$ of the twin spacing $h$,
	\begin{equation}
		\label{eq:phielast}
		\frac{E_{\rm elast}^{(i)}}{2 h_i L_i} = \frac12 A (h_{,x})^2 , \qquad
		A = \frac{9}{16} \mu \lambda (1-\lambda)^3 \alpha^2 g_2 ,
	\end{equation}
and this estimate will be used in our model.

Note that the factor $\frac23$ in Eq.~\eqref{eq:dhdx} results from the specific construction adopted here, see Fig.~\ref{fig:Seiner}(c), and is considered a reasonable estimate. However, factors other than $\frac23$ could be adopted in the range between $\frac12$ and $1$, as also illustrated in Fig.~\ref{fig:Seiner}(c).

\subsection{Governing equations}
\label{sec:governing}

The governing equations of the model will now be derived through minimization of the total free energy of the system. 
The free energy density (per unit volume of the branched microstructure) is formulated in the following form,
	\begin{equation}
		\label{eq:phi}
		\phi(h,h_{,x}) = \frac12 A (h_{,x})^2 + \frac{2\gamma}{\alpha h} .
	\end{equation}
The first term is the elastic strain energy contribution derived above, see Eq.~\eqref{eq:phielast}. 
The second term describes the interfacial energy of twin boundaries and is a continuous counterpart of Eq.~\eqref{eq:Eint}${}_2$. 
The free energy function $\phi$ is the density per unit volume, i.e., the density per unit area of the A--MM interface and per unit length along the $x$-axis. 

The total free energy functional (density per unit area of the A--MM interface) is now obtained by integrating the free energy density $\phi$ over the length of the domain,
	\begin{equation}
		\label{eq:Phi}
		\Phi = \int_0^L \phi(h,h_{,x}) \rd x + \Gamma \, h \big\vert_{x=L} .
	\end{equation}
The total free energy includes also the energy of elastic micro-strains at the A--MM interface, which is introduced by the second term in Eq.~\eqref{eq:Phi}. It can be shown that this energy is proportional to the twin spacing at $x=L$, denoted by $h\big\vert_{x=L}$, as indicated in Eq.~\eqref{eq:Phi}, and $\Gamma$ is the corresponding proportionality factor. 
\citet{seiner2020branching} have developed a separate construction to estimate $\Gamma$, see their Fig.~7(b). 
An alternative approach is to estimate $\Gamma$ by solving an adequate boundary value problem combined with shape optimization, which can be performed using either a sharp-interface framework \citep{maciejewski2005elastic,stupkiewicz2007low} or a diffuse-interface framework within the phase-field method \citep{tuuma2016phase,amini2023energy}.
Below, we adopt the value of $\Gamma$ by referring to the results of phase-field simulations, as discussed in Section~\ref{sec:predictions}.

The total free energy $\Phi$ involves four parameters, namely $\gamma$, $A$, $\alpha$, and $\Gamma$, along with the size $L$ of the branched domain. The twin boundary energy $\gamma$ is a material parameter. Parameters $A$, $\alpha$, and $\Gamma$ depend on the specific A--MM microstructure. Moreover, $A$ and $\Gamma$ depend on the elastic properties of the phases.

We assume that the microstructure of the branched martensite domain is formed such that the total free energy is minimized. This is a common approach in the modeling of martensitic microstructures, followed in particular in the crystallographic theory of martensite \citep{ball1987fine,Bhattacharya2003}, as well as in the modeling of twin branching \citep{kohn1992branching,seiner2020branching}. 
This assumption will be relaxed in Section~\ref{sec:dissip} where the evolution of the microstructure will be considered and the associated energy dissipation will be accounted for.
The twin spacing $h(x)$ is thus found by minimizing the total free energy $\Phi$,
	\begin{equation}
		\Phi \;\to\; \min_h .
	\end{equation}
The necessary condition for the minimum of $\Phi$, i.e.\ the condition of stationarity of $\Phi$ with respect to an arbitrary variation $\delta h$ of $h$, reads
	\begin{equation}
		\label{eq:weak}
		0 = \delta \Phi = \int_0^L \left( A h_{,x} \delta h_{,x} - \frac{2\gamma}{\alpha h^2} \delta h \right) \rd x 
		+ \Gamma \, \delta h \big\vert_{x=L}
		\qquad \forall \, \delta h .
	\end{equation}

The local (strong) form of the governing equation can then be obtained in a standard manner. 
Using the divergence theorem, Eq.~\eqref{eq:weak} is transformed to
	\begin{equation}
		\label{eq:weak2}
		\int_0^L \left( -A h_{,xx} - \frac{2\gamma}{\alpha h^2} \right) \delta h \, \rd x 
		- A h_{,x} \delta h \big\vert_{x=0} + \big( A h_{,x} + \Gamma \big) \delta h \big\vert_{x=L} = 0
		\qquad \forall \, \delta h ,
	\end{equation}
where the whole second term is evaluated at $x=0$ (i.e., both $h_{,x}$ and $\delta h$, as denoted by the vertical bar with a subscript), and likewise the whole third term is evaluated at $x=L$.
Following the standard argument, Eq.~\eqref{eq:weak2} implies the following local form of the governing equation,
	\begin{equation}
		\label{eq:strong}
		h_{,xx} + \frac{2\gamma}{\alpha A h^2} = 0 \quad \mbox{for} \; x \in [0,L] ,
	\end{equation}
along with the boundary conditions
	\begin{equation}
		\label{eq:strong:bc}
		h_{,x} \big\vert_{x=0} = 0 , \qquad
		h_{,x} \big\vert_{x=L} = -\frac{\Gamma}{A} .
	\end{equation}
Note that the boundary condition at $x=0$, Eq.~\eqref{eq:strong:bc}${}_1$, is a continuous counterpart of the non-branched first segment in the construction proposed by \citet{seiner2020branching}, see Fig.~\ref{fig:Seiner}(a).

Let us multiply Eq.~\eqref{eq:strong} by $h_{,x}$ (and by $A$) and integrate it once over $x$. As a result, the following property of the solution of the problem at hand is obtained,
	\begin{equation}
		\label{eq:equipart}
		\frac12 A (h_{,x})^2 - \frac{2\gamma}{\alpha h} = C ,
	\end{equation}
where $C$ is an integration constant. 
This is the property of \emph{equipartition of energy} \citep{kohn1992branching}, which shows that the difference between the elastic strain energy $\frac12 A (h_{,x})^2$ and the interfacial energy $2\gamma/(\alpha h)$ is \emph{independent} of $x$.

Introducing dimensionless twin spacing $\bar{h}$ and dimensionless coordinate $\bar{x}$, both normalized using $L$,
	\begin{equation}
		\label{eq:hbar}
		\bar{h} = \frac{h}{L} , \qquad
		\bar{x} = \frac{x}{L} ,
	\end{equation}
Eqs.~\eqref{eq:strong} and~\eqref{eq:strong:bc} can be written in a dimensionless form, i.e.,
	\begin{equation}
		\label{eq:strong:dimless}
		\bar{h}_{,\bar{x}\bar{x}} + \frac{2\bar{\gamma}}{\bar{h}^2} = 0 \quad \mbox{for} \; \bar{x} \in [0,1] , \qquad
		\bar{h}_{,\bar{x}} \big\vert_{\bar{x}=0} = 0 , \qquad
		\bar{h}_{,\bar{x}} \big\vert_{\bar{x}=1} = -\bar{\Gamma} .
	\end{equation}
Accordingly, the model involves two dimensionless material parameters,
	\begin{equation}
		\label{eq:gammabar}
		\bar{\gamma} = \frac{\gamma}{\alpha A L} , \qquad
		\bar{\Gamma} = \frac{\Gamma}{A} .
	\end{equation}
For future use, let us also introduce dimensionless free energy density $\bar{\phi}$ and dimensionless total free energy $\bar{\Phi}$,
	\begin{equation}
		\label{eq:phi-bar}
		\bar{\phi} = \frac{\phi}{A} = \frac12 (\bar{h}_{,\bar{x}})^2 + \frac{2\bar{\gamma}}{\bar{h}} , \qquad
		\bar{\Phi} = \frac{\Phi}{AL}
		  = \int_0^1 \bar{\phi}(\bar{h},\bar{h}_{,\bar{x}}) \rd\bar{x} + \bar{\Gamma} \, \bar{h} \big\vert_{\bar{x}=1} .
	\end{equation}
The dimensionless counterpart of the weak form~\eqref{eq:weak} reads
	\begin{equation}
		\label{eq:weak:dimless}
		\int_0^1 \left( \bar{h}_{,\bar{x}} \delta \bar{h}_{,\bar{x}} - \frac{2\bar{\gamma}}{\bar{h}^2} \delta \bar{h} \right) \rd \bar{x} 
		+ \bar{\Gamma} \, \delta \bar{h} \big\vert_{\bar{x}=1} = 0
		\qquad \forall \, \delta \bar{h} .
	\end{equation}
The above weak form will be used as a basis for an approximate solution of the problem using the finite element method.
Note that the problem at hand admits an analytical solution that is derived in \ref{app:analytical}. However, the solution requires solving two nonlinear equations numerically. Since the finite element method is indispensable for the evolution problem studied in Section~\ref{sec:dissip}, it is also used for this static case.

\subsection{Finite-element treatment}
\label{sec:FE}

While the weak form~\eqref{eq:weak} and its dimensionless counterpart~\eqref{eq:weak:dimless} are fully equivalent, the latter appears to be more convenient in computational practice, as follows from our preliminary simulations. 
The finite-element solution is thus obtained by introducing a standard piecewise-linear approximation of the dimensionless twin spacing $\bar{h}$ over the interval $[0,1]$. 
Application of the Galerkin method delivers then a set of nonlinear equations that are solved using the Newton method. 
Note that an adequate initialization of the iterative scheme is needed to avoid division by zero in view of the term $\bar{h}^2$ in the denominator in the weak form~\eqref{eq:weak:dimless}. 
Further, the nodal (Lobatto) integration is employed, since convergence problems or non-physical solutions with $\bar{h}<0$ are often encountered in the case of the Gauss integration.
Standard details of the finite-element treatment are omitted here.

Considering that the actual solutions may be characterized by a high gradient in the vicinity of $\bar{x}=1$, see e.g.\ Fig.~\ref{fig:sample}(a) below, the finite-element mesh is significantly refined towards $\bar{x}=1$. 
As a result, highly accurate solutions can be obtained in most cases for the number of elements of the order of 100. However, in the most demanding cases ($\bar{\gamma}<10^{-6}$), a much finer mesh of $10^4$ elements is needed to obtain a converged solution.

\subsection{Model predictions}
\label{sec:predictions}

\begin{figure}
\vspace{2ex}
\centerline{
	\begin{tabular}{rrrr}
		{\footnotesize (a)}\!\!\! &
		& ~~{\footnotesize (b)}\!\!\! & \\[-2.5ex]
		& \inclps{!}{0.37\textwidth}{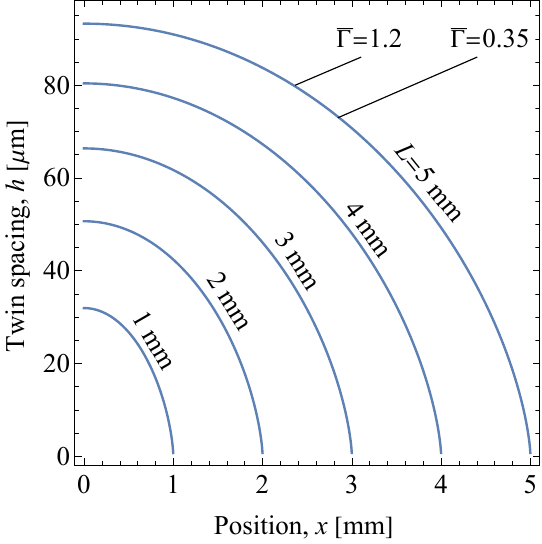} &
		& \inclps{!}{0.37\textwidth}{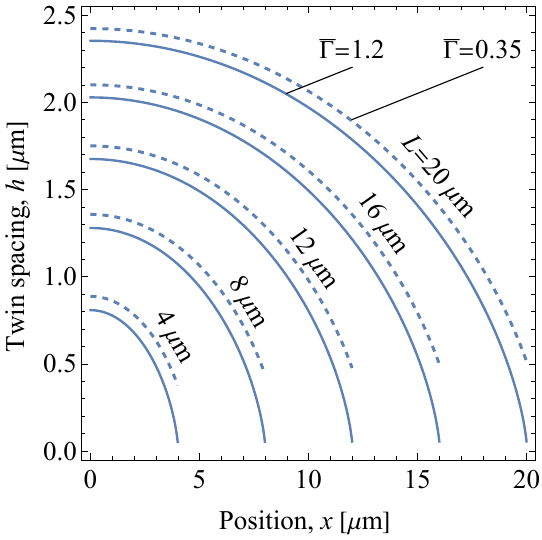}
	\end{tabular}
	}
\caption{Twin spacing $h(x)$ for (a) relatively large domain with $L$ varied between 1\,mm ($\bar{\gamma}=2\times10^{-5}$) and 5\,mm ($\bar{\gamma}=4\times10^{-6}$) and (b) relatively small domain with $L$ varied between 4\,$\mu$m ($\bar{\gamma}=5\times10^{-3}$) and 20\,$\mu$m ($\bar{\gamma}=10^{-3}$). In panel (a), the solid lines ($\bar{\Gamma}=1.2$) and the dashed lines ($\bar{\Gamma}=0.35$) overlap.}
\label{fig:sample}
\end{figure}

Sample results are shown in Fig.~\ref{fig:sample} in order to illustrate the predictions delivered by the model. 
The material parameters correspond to those adopted by \citet{seiner2020branching} for the CuAlNi alloy, and the same type of A--MM microstructure is considered. Specifically, the twin boundary energy is adopted as $\gamma=0.1$\,J/m${}^2$, and parameter $A$ characterizing the elastic strain energy of branching is computed according to Eq.~\eqref{eq:phielast}${}_2$ using $\mu=70$\,GPa, $\lambda=0.6992$, $g_2=0.03648$, and $\alpha=0.5642$, so that $A=8.7$\,MJ/m${}^3$. 

The last model parameter to be determined is the elastic micro-strain energy factor $\Gamma$. 
In their supplementary material, \citet{seiner2020branching} provide an upper-bound estimate of the elastic energy of closure domains that could be used to compute $\Gamma$. Instead of following this route, we identify $\Gamma$ from the results reported in their Fig.~8(a). Specifically, for $N=0$, i.e., for the case with no branching, the total energy in the twinned martensite, denote it by $\Phi^\ast$ ($E^{(0)}$ in their notation), has been reported to be equal to 270\,N/m. On the other hand, for a constant $h$, the total energy is given by $\Phi^\ast=2\gamma L/(\alpha h)+\Gamma h$, which upon minimization with respect to $h$ is found equal to $\Phi^\ast=\sqrt{8L\Gamma\gamma/\alpha}$. Finally, parameter $\Gamma=10.3$\,MJ/m${}^3$ can be computed from the latter formula. The corresponding dimensionless parameter is equal to $\bar{\Gamma}=1.2$, see Eq.~\eqref{eq:gammabar}${}_2$.

As a reference, another value of parameter $\Gamma$ will also be used in the computations. This will, in particular, illustrate the sensitivity of the results to the value of $\Gamma$. 
To this end, parameter $\Gamma$ will be estimated based on the results of the micromechanical analysis carried out using the phase-field method \citep{tuuma2016phase}.
Out of the four crystallographically distinct microstructures of the A--MM interface, the one considered by \citet{seiner2020branching} corresponds to microstructure M3 in the notation used by \citet{tuuma2016phase}. 
The corresponding values of the energy factor $\Gamma^{\rm e}_{\rm am}$ reported in Fig.~10 in \citet{tuuma2016phase} range from about 2\,MJ/m${}^3$ to 5\,MJ/m${}^3$ depending on the twin spacing and interface thickness parameter of the phase-field model. 
The lower values of $\Gamma^{\rm e}_{\rm am}$ are associated with a local branching mechanism that becomes effective when the twin spacing is greater than 30\,nm, see Fig.~6 in \citet{tuuma2016phase}, and also with a greater interface thickness in the phase-field model. 
Therefore, we adopt the value of $\Gamma^{\rm e}_{\rm am}\approx5$\,MJ/m${}^3$ as a reasonable estimate, which is also consistent with the results of the sharp-interface modeling reported by \citet{stupkiewicz2007low}. 
Note that the factor $\Gamma^{\rm e}_{\rm am}$ in \citet{tuuma2016phase} corresponds to the twin spacing measured in the direction normal to the twin interfaces, while the factor $\Gamma$ in Eq.~\eqref{eq:Phi} corresponds to the twin spacing measured along the A--MM interface. Accordingly, we have $\Gamma=\alpha\Gamma^{\rm e}_{\rm am}$, and thus the value of $\Gamma=3$\,MJ/m${}^3$ will be used in the following (the corresponding dimensionless parameter is equal to $\bar{\Gamma}=0.35$). 
This value is three times lower than that used by \citet{seiner2020branching}, and we believe it is a more accurate estimate of $\Gamma$ than their upper-bound estimate, as it is determined by a kind of shape optimization delivered by the phase-field method and it satisfies the continuum equilibrium equations, in contrast to the upper-bound estimate of \citet{seiner2020branching}.

Finally, two ranges of the size $L$ of the twinned martensite domain are considered. 
The first range corresponds to a relatively large domain with $L$ varied between $1$\,mm and $5$\,mm, where the latter value was used in the main illustrating example of \citet{seiner2020branching}, see their Fig.~8. 
The respective results are shown in Fig.~\ref{fig:sample}(a). 
The second range corresponds to a relatively small domain with $L$ varied between $4\,\mu$m and $20\,\mu$m, and the respective results are shown in Fig.~\ref{fig:sample}(b). 
It can be seen that twin branching and the corresponding refinement of the microstructure towards the A--MM interface is more pronounced for larger domains. 
The effect of parameter $\bar{\Gamma}$ characterizing the elastic micro-strain energy at the A--MM interface is negligible in the case of relatively large domain, as shown in Fig.~\ref{fig:sample}(a), except in the vicinity of the A--MM interface, however, the corresponding difference is not discernible in Fig.~\ref{fig:sample}(a). In the case of the relatively small domain, Fig.~\ref{fig:sample}(b), the effect of $\bar{\Gamma}$ is visible, and it is more pronounced at the A--MM interface, at $x=L$, than at the free end, at $x=0$.

\begin{figure}
\centerline{
	\begin{tabular}{rrrr}
		{\footnotesize (a)}\!\!\! &
		& ~~{\footnotesize (b)}\!\!\! & \\[-4.5ex]
		& \inclps{!}{0.37\textwidth}{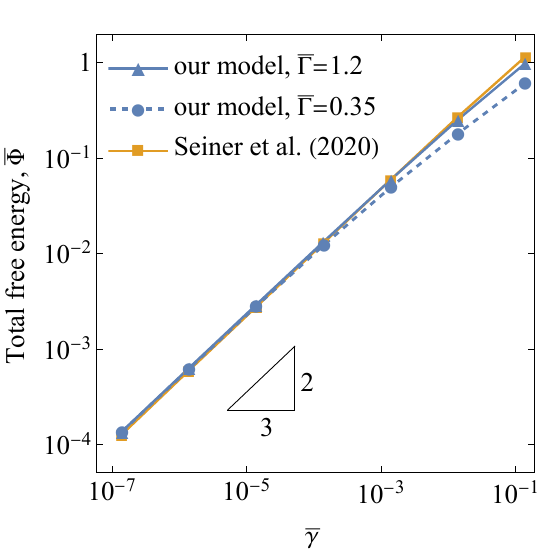} &
		& \inclps{!}{0.37\textwidth}{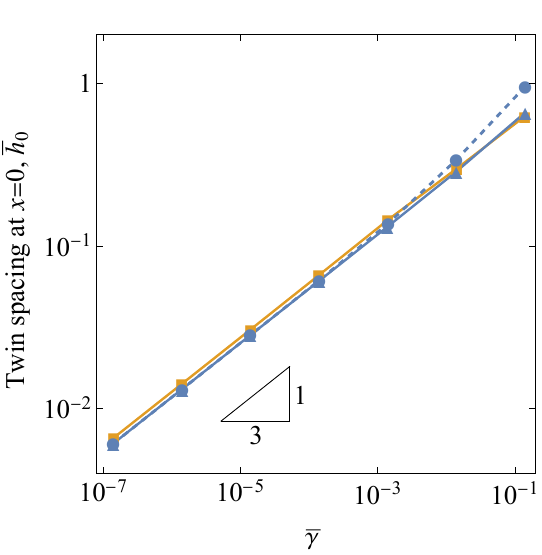} \\
		{\footnotesize (c)}\!\!\! &
		& ~~{\footnotesize (d)}\!\!\! & \\[-4.5ex]
		& \inclps{!}{0.37\textwidth}{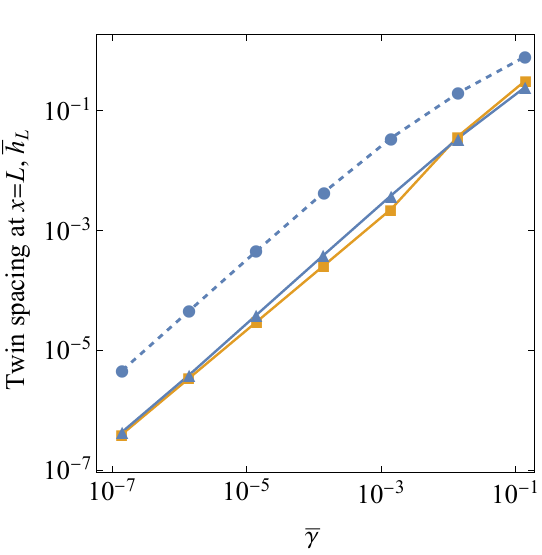} &
		& \inclps{!}{0.37\textwidth}{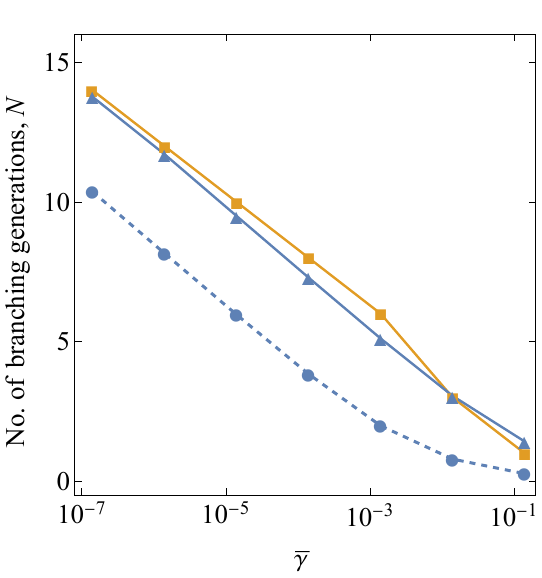}
	\end{tabular}
	}
\caption{Predictions of the present continuous model compared to those of the discrete model of \citet{seiner2020branching}: (a) dimensionless total free energy $\bar{\Phi}$, (b) dimensionless twin spacing $\bar{h}_0$ at $x=0$, (c) dimensionless twin spacing $\bar{h}_L$ at $x=L$, and (d) the number of branching generations $N$; all as a function of the dimensionless twin boundary energy $\bar{\gamma}$, see Eq.~\eqref{eq:gammabar}${}_1$.}
\label{fig:compare}
\end{figure}

Fig.~\ref{fig:compare} presents a comprehensive comparison of the predictions of the present continuous model with those of the discrete model of \citet{seiner2020branching}, as reported in their Table~1. 
Their results have been reported for the length $L$ varied between $10^4$\,nm and $10^7$\,nm (four values) and for the material length $\gamma/\mu$ varied between $10^{-4}$\,nm and $10^{-1}$\,nm (four values). Apparently, the 16 cases reported by \citet{seiner2020branching} collapse with a very good accuracy to seven distinct cases in our dimensionless representation, Eqs.~\eqref{eq:hbar}, \eqref{eq:gammabar} and \eqref{eq:phi-bar}, with the dimensionless twin boundary energy $\bar{\gamma}$ spanning six decades from about $10^{-7}$ to $10^{-1}$.

Fig.~\ref{fig:compare} shows the dimensionless total free energy $\bar{\Phi}$ as a function of $\bar{\gamma}$ and the dimensionless twin spacing at the free end ($x=0$) and at the A--MM interface ($x=L$), denoted by $\bar{h}_0$ and $\bar{h}_L$, respectively. Moreover, the number of branching generations $N$ is shown in Fig.~\ref{fig:compare}(d). In the discrete model of \citet{seiner2020branching}, $N$ is a positive integer such that $h_L=h_N=h_0/2^N$. 
In the case of our continuous model, an analogous quantity $N$ can be defined,
	\begin{equation}
		N=\log_2 \left( \frac{h_0}{h_L} \right) ,
	\end{equation}
which, clearly, takes non-integer values. Since branching is assumed to proceed in a continuous manner, rather than in the form of distinct branching generations, this number is meant to quantify the microstructure refinement, and does not indicate the actual number of distinct branching generations.

As discussed above, the computations have been carried out for two values of parameter $\bar{\Gamma}$. An almost perfect agreement with the results of \citet{seiner2020branching} is obtained in the entire range of $\bar{\gamma}$ when $\bar{\Gamma}=1.2$, i.e., when the same input parameters are used in the two models, see Fig.~\ref{fig:compare}. This is a remarkable result, in particular, for the largest values of $\bar{\gamma}$ for which only one branching generation is predicted by the model of \citet{seiner2020branching}, and the major characteristics of the solution are correctly represented by our continuous model.

The results corresponding to a lower value of $\bar{\Gamma}=0.35$ are also included in Fig.~\ref{fig:compare}, see the dashed lines, and illustrate the sensitivity of the results to $\bar{\Gamma}$. It follows that the total energy $\bar{\Phi}$ and the twin spacing $\bar{h}_0$ at the free end are only weakly affected by $\bar{\Gamma}$ for the largest values of $\bar{\gamma}$ and are practically insensitive to $\bar{\Gamma}$ for $\bar{\gamma}$ varied between $10^{-7}$ and $10^{-3}$. 
The twin spacing at the A--MM interface, $\bar{h}_L$, is visibly affected by $\bar{\Gamma}$, and this is expected since parameter $\bar{\Gamma}$ governs the elastic micro-strain energy at the A--MM interface and hence directly impacts $\bar{h}_L$. As a result, also the number of twin branching generations $N$ is visibly affected by $\bar{\Gamma}$. For $\bar{\gamma}=10^{-1}$ (the largest value of $\bar{\gamma}$ considered), the continuous model predicts $N=0.27$ with the interpretation that only a fraction of twin lamellae would branch (the discrete model predicts then one branching generation, $N=1$).

The theory of \citet{kohn1992branching} predicts that, in the case of the self-similar construction, i.e., for small $\gamma$ and large $\Gamma$, the total free energy $\Phi$ and the twin spacing at the free end, $h_0$, scale according to $\Phi\sim\mu^{1/3}\gamma^{2/3}L^{1/3}$ and $h_0\sim\mu^{-1/3}\gamma^{1/3}L^{2/3}$, respectively. \citet{seiner2020branching} have checked that their results also satisfy these scaling laws with a good accuracy, see their Fig.~8(d,e). 
Now, in our dimensionless representation, the scaling laws of \citet{kohn1992branching} take a particularly simple form, namely
    \begin{equation}
        \bar{\Phi}\sim\bar{\gamma}^{2/3} , \qquad
        \bar{h}_0\sim\bar{\gamma}^{1/3} .
    \end{equation}
As can be seen from Fig.~\ref{fig:compare}(a,b), the results of \citet{seiner2020branching} indeed follow these two scaling laws. 
Apparently, it is also the case of our results for $\bar{\Gamma}=1.2$. 
However, for $\bar{\Gamma}=0.35$, our results deviate from the two scaling laws for $\bar{\gamma}>10^{-3}$, thus indicating a deviation from self-similar regime of \citet{kohn1992branching}.

\begin{figure}
\centerline{
	\begin{tabular}{rrrr}
		{\footnotesize (a)}\!\!\! &
		& ~~{\footnotesize (b)}\!\!\! & \\[-4.5ex]
		& \inclps{!}{0.37\textwidth}{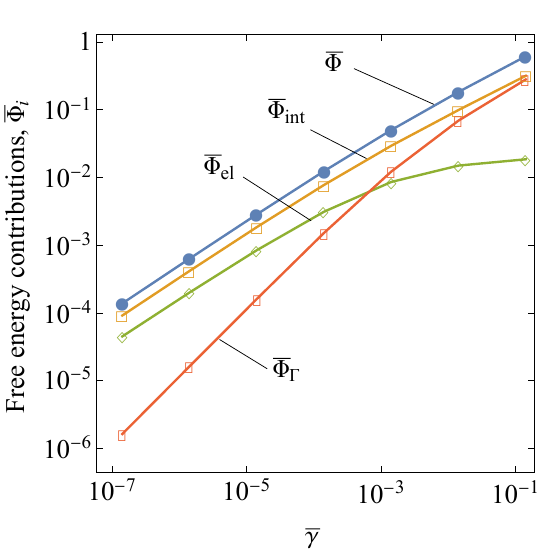} &
		& \inclps{!}{0.37\textwidth}{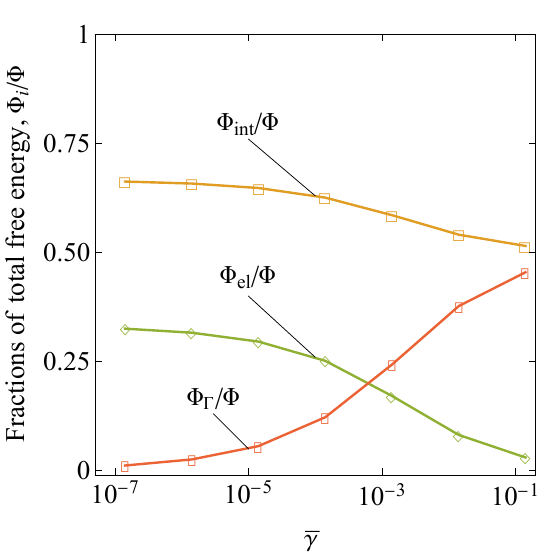}
	\end{tabular}
	}
\caption{Individual contributions to the total free energy $\Phi$ as a function of $\bar{\gamma}$ (for $\bar{\Gamma}=0.35$): (a) dimensionless energy contributions $\bar{\Phi}_i$; (b) fractions $\Phi_i/\Phi$.}
\label{fig:contributions}
\end{figure}

Fig.~\ref{fig:contributions} showcases the individual contributions $\Phi_k$ to the total free energy $\Phi$ as a function of $\bar{\gamma}$, where
	\begin{equation}
		\label{eq:Phi:contrib}
		\Phi_{\rm el} = \int_0^L \frac12 A (h_{,x})^2 \rd x , \qquad
		\Phi_{\rm int} = \int_0^L \frac{2\gamma}{\alpha h} \rd x , \qquad
		\Phi_\Gamma = \Gamma \, h \big\vert_{x=L} ,
	\end{equation}
see Eqs.~\eqref{eq:phi} and~\eqref{eq:Phi}. 
The interfacial energy contribution $\Phi_{\rm int}$ constitutes the largest fraction of $\Phi$ over the entire range of $\bar{\gamma}$. 
The contribution of the elastic strain energy, $\Phi_{\rm el}$, amounts to about one third of $\Phi$ for small $\bar{\gamma}$ and decreases towards zero for $\bar{\gamma}=10^{-1}$. This naturally correlates with the intensity of twin branching characterized by $N$, see Fig.~\ref{fig:compare}(d). 
Finally, the contribution of the elastic micro-strain energy at the A--MM interface, $\Phi_\Gamma$, is negligible for $\bar{\gamma}=10^{-7}$ and reaches nearly 50\% for $\bar{\gamma}=10^{-1}$. In this latter regime, where $\bar{\gamma}$ is large and $N$ tends to zero (no branching), the two contributions $\Phi_{\rm int}$ and $\Phi_\Gamma$ tend to be equal. Equality of the two contributions is a result that is consistent with a simple model minimizing the total interfacial energy of a twinned plate in the absence of twin branching \citep{petryk2006modeling,petryk2010interfacial:part2}.

\section{Introducing dissipation effects}
\label{sec:dissip}

\subsection{Overall dissipation potential}

Evolution of the branched microstructure is associated with propagation of twin interfaces, and this propagation is now considered to be a dissipative process. Below, the constitutive description of the dissipative behavior is introduced at the local level of twin interfaces, and then it is upscaled to the macroscopic level by referring to the specific branched microstructure shown in Fig.~\ref{fig:Seiner}. 

In the continuous setting, the twin spacing $h$ is {thus} now considered as a function of both position and time, $h=h(x,t)$. Additionally, the length of the twinned domain is time-dependent, such that $0\leq x\leq L(t)$. 

{Note that the length $L(t)$ is here adopted as a prescribed control parameter. Accordingly, the dissipation associated with the propagation of the A--MM interface does not intervene in the model. This dissipation would affect the evolution of $L$ as a function of some external stimuli, but this is not considered in the model. In principle, this dissipation can affect the evolution of the local microstructure of the direct A--MM interface, which, in turn, can influence the value of the energy factor $\Gamma$. This is expected to be a secondary effect (if any), and it is not considered in the model, hence $\Gamma$ is assumed to be constant.}

{Considering thus the propagation of twin interfaces, we} assume that, at the local level, the dissipative behavior is governed by a local rate-independent dissipation potential (density per unit interface area) that depends on the local interface propagation speed $\hat{v}_{\rm n}$ (normal to the interface),
	\begin{equation}
		\label{eq:Dri}
		\hat{D} = \hat{r} | \hat{v}_{\rm n} | ,
	\end{equation}
where $\hat{r}$ is the rate-independent threshold (critical driving force) for interface propagation, and the quantities with a hat refer to the local sharp-interface description. 
The dependence of the interface speed $\hat{v}_{\rm n}$ on the rate of change of the twin spacing, $\dot{h}$, is derived in~\ref{app:speed}, see Eq.~\eqref{eq:vn2}, and reads
	\begin{equation}
		\label{eq:vn}
		\hat{v}_{\rm n} = \frac34 (1-\lambda) \alpha \dot{h} .
	\end{equation}

Referring to the discrete setting, the total dissipation potential $D^{(i)}$ (density per unit volume) for a single cell, see Fig.~\ref{fig:Seiner}(b), is obtained by averaging over the cell volume. Considering that within the cell there are four interfaces of the length $L_i/\alpha$ (for $L_i\gg h_i$) and assuming that the interface speed $\hat{v}_{\rm n}$ is identical for all interfaces, we have
	\begin{equation}
		D^{(i)} = \frac{1}{2 h_i L_i} \frac{4 L_i}{\alpha} \hat{D} = \frac{2 \hat{D}}{\alpha h_i} ,
	\end{equation}
and in the continuous setting
	\begin{equation}
		\label{eq:D}
		D = \frac{2 \hat{D}}{\alpha h} .
	\end{equation}
Combining Eqs.~\eqref{eq:Dri}, \eqref{eq:vn}, and~\eqref{eq:D}, the overall dissipation potential is finally obtained in the following form,
	\begin{equation}
		\label{eq:D:hdot}
		D = \frac{r}{h} | \dot{h} | , \qquad
		r = \frac32 (1-\lambda) \hat{r} .
	\end{equation}
We note that $D$ depends on $|\dot{h}|$ just like $\hat{D}$ depends on $|v_{\rm n}|$, hence the structure of the dissipation potential is preserved, while the effective threshold is equal to $r/h$ and is inversely proportional to~$h$.

The above simple description of the dissipative effects is based on the assumption that the dissipation results from the propagation of twin interfaces only. 
However, the actual evolution of the branched microstructure may also be associated with abrupt evens, like formation or disappearance of individual twin lamellae. In agreement with the point of view adopted in the present continuous model, such events are assumed to appear locally at different locations and at different time instants, hence are not observed on the macroscale upon spatial and time averaging. Secondly, the energy dissipation associated with such abrupt events may require a different description than that corresponding to a smooth propagation of twin interfaces. In the present model, such effects can only be accounted for by adjusting the rate-independent threshold $\hat{r}$. A more refined description would be an interesting topic for future work.

\subsection{Evolution problem}
\label{sec:evolution}

Consider now the evolution problem in which the length $L=L(t)$ of the twinned domain is prescribed as a function of time, where $L(t)$ plays the role of a control parameter. 
To determine the rate of change of the twin spacing, $\dot{h}$, we formulate a minimization problem for the rate potential $\Pi$,
	\begin{equation}
		\label{eq:min}
		\Pi = \dot{\Phi} + {\cal D} \;\to\; \min_{\dot{h}} ,
	\end{equation}
where $\dot{\Phi}$ is the rate of change of the total free energy,
	\begin{equation}
		\dot{\Phi} = \int_0^L \left( A h_{,x} \dot{h}_{,x} - \frac{2\gamma}{\alpha h^2} \dot{h} \right) \rd x 
		+ \Gamma \, \dot{h} \big\vert_{x=L}
		+ \dot{L} \phi \big\vert_{x=L} ,
	\end{equation}
and ${\cal D}$ is the total dissipation potential,
	\begin{equation}
		\label{eq:Dtotal}
		{\cal D} = \int_0^L D \, \rd x .
	\end{equation}

The necessary condition for the minimum of $\Pi$ reads
	\begin{equation}
		\label{eq:weak:ri}
		0 \in \int_0^L \left( A h_{,x} \delta\dot{h}_{,x} - \frac{2\gamma}{\alpha h^2} \delta\dot{h} +
		\partial_{\dot{h}} | \dot{h} | \, \frac{r}{h} \delta\dot{h} \right) \rd x 
		+ \Gamma \, \delta\dot{h} \big\vert_{x=L}
		 \qquad \forall \, \delta \dot{h} ,
	\end{equation}
where the subdifferential $\partial_{\dot{h}}|\dot{h}|$ and the resulting inclusion are introduced in view of the non-differentiability of the dissipation potential.
Due to the non-smoothness, a direct computational treatment of the above minimization problem is not immediate, and the augmented Lagrangian method \citep{alart1991mixed} is used for that purpose. The details are provided in \ref{app:AL}.

Let us now derive a dimensionless form of Eq.~\eqref{eq:weak:ri}. As in Section~\ref{sec:governing}, the twin spacing $h$ is normalized by the domain length $L$, thus $\bar{h}=h/L$, and the dimensionless twin spacing $\bar{h}=\bar{h}(\bar{x},t)$ is defined on a fixed domain, $0\leq\bar{x}\leq1$. Since the evolution problem is rate-independent, the solution is invariant upon arbitrary scaling of the time, and hence the time $t$ is not normalized. 

The time derivative of the twin spacing, $\dot{h}=\partial h/\partial t$, enters the weak form~\eqref{eq:weak:ri} through the subdifferential $\partial_{\dot{h}}|\dot{h}|$. 
To maintain the physical meaning of $\dot{h}$ in the dimensionless setting, the time derivative of the dimensionless twin spacing $\bar{h}$ must be taken at a constant $x$ rather than at a constant $\bar{x}$. The corresponding time derivative is denoted by $\overset{\mbox{\tiny $\circ$}}{\bar{h}}$ and satisfies
	\begin{equation}
		\label{eq:hcirc}
		\overset{\mbox{\tiny $\circ$}}{\bar{h}} (\bar{x},t) = \frac1L \, \dot{h} (x,t) \big\vert_{x=L\bar{x}} ,
	\end{equation}
see \ref{app:AL} for the computational treatment of this term.

The weak form~\eqref{eq:weak:ri} involves also the test function $\delta\dot{h}$ that has an interpretation of the variation of the twin spacing rate. 
It is an arbitrary function of sufficient regularity, just like $\delta h$ in Eq.~\eqref{eq:weak}. Accordingly, the test function $\delta\dot{h}$ can be equivalently replaced by $\delta h$ in the weak form~\eqref{eq:weak:ri}, and subsequently $\delta\bar{h}=\delta h/L$ will be used as the test function in the dimensionless formulation of the weak form~\eqref{eq:weak:ri} which reads
	\begin{equation}
		\label{eq:weak:ri:dl}
		0 \in \int_0^1 \left( \bar{h}_{,\bar{x}} \delta\bar{h}_{,\bar{x}} - \frac{2\bar{\gamma}}{\bar{h}^2} \delta\bar{h} +
		\partial_{\overset{\mbox{\tiny $\circ$}}{\bar{h}}} \big| \overset{\mbox{\tiny $\circ$}}{\bar{h}} \big| \, \frac{\bar{r}}{\bar{h}} \delta\bar{h} \right) \rd \bar{x} 
		+ \bar{\Gamma} \, \delta\bar{h} \big\vert_{\bar{x}=1}
		 \qquad \forall \, \delta \bar{h} .
	\end{equation}
Compared to the static case in Eq.~\eqref{eq:weak}, the above weak form of the evolution problem involves an extra term introducing the dissipative driving force, where $\bar{r}/\bar{h}$ is the dimensionless threshold for interface propagation with
	\begin{equation}
		\label{eq:rbar}
		\bar{r} = \frac{r}{A} = \frac{3(1-\lambda)\hat{r}}{2A} .
	\end{equation}

The weak form~\eqref{eq:weak:ri:dl} constitutes the basis for the finite-element treatment of the evolution problem. Despite the domain $0\leq x\leq L$ evolves in time, the actual computations are carried out on a fixed domain, $0\leq\bar{x}\leq1$, and on a fixed finite-element mesh. This simplifies enforcement of the boundary condition at $x=L$ (at $\bar{x}=1$), but requires a proper computation of the rate of the twin spacing. 
This is discussed further in \ref{app:AL}.

\subsection{Model predictions}

The effect of energy dissipation on the evolution of twin spacing $h$ in the branched microstructure is studied in this section. We consider the scenario in which the length of the twinned domain $L$ increases from $L_0$ to $L_{\rm max}$, followed by a decreases back to $L_0$. 
We set the initial length as $L_0=0.01L_{\rm max}$. 
Since the evolution is here rate-independent, the length $L$ is a control parameter and plays the role of a pseudo-time.
The initial twin spacing corresponding to $L=L_0$ is determined by minimization of the total free energy, as in Section~\ref{sec:predictions}. 

The evolution problem at hand involves one additional material parameter to be specified, namely the rate-independent threshold $\hat{r}$, see Eq.~\eqref{eq:Dri}. The remaining parameters are the same as in the study in Section~\ref{sec:predictions}. Note that $\Gamma=3$\,MJ/m${}^3$ is adopted here. The rate-independent threshold is expected to strongly depend on the material, its state (purity, defects, etc.), and on the twinning mode (e.g., type I or type II twinning). To estimate the range of physically-relevant values of $\hat{r}$, we consider the work of \citet{novak2006transformation} who performed a comprehensive study on martensite variant reorientation in prism-shaped single crystal of CuAlNi alloy. In the experiment, one variant of martensite is transformed to another variant by compressing the specimen along a specified direction. The transformation proceeds through the evolution of a simple laminate composed of the two variants. Using the respective stress--strain curve, the critical driving force for transformation, which at the microscale corresponds to the rate-independent threshold for interface propagation, can be determined. Accordingly, we have $\hat{r}=\sigma^{\rm tw}\varepsilon^{\rm tw}$, where $\sigma^{\rm tw}$ and $\varepsilon^{\rm tw}$ denote the uniaxial transformation (twinning) stress and strain, respectively. It follows that, for type II twinning, the parameter $\hat{r}$ is within the range between 1.25\,MPa ($\sigma^{\rm tw}=25$\,MPa, $\varepsilon^{\rm tw}=0.05$, see Fig.~3b in~\citet{novak2006transformation}) and 4\,MPa ($\sigma^{\rm tw}=130$\,MPa, $\varepsilon^{\rm tw}=0.033$, see Fig.~3c in~\citet{novak2006transformation}). 
For compound twins, the threshold is much lower, $\hat{r}\approx0.03$\,MPa ($\sigma^{\rm tw}\approx1$\,MPa, $\varepsilon^{\rm tw}=0.03$, see the V2$\to$V1 transformation in Fig.~2b in~\citet{novak2006transformation}). Finally, it is worthwhile to note that type I twins have not been observed in their study. 
In the following analysis, we will use a relatively high value of $\hat{r}=4$\,MPa, as determined above, so that the effects of energy dissipation are well visible. 
According to Eq.~\eqref{eq:rbar}, the corresponding dimensionless parameter is here $\bar{r}=0.21$.

For the purpose of presentation of the results, it is convenient to introduce an alternative normalization based on the maximum length $L_{\rm max}$, rather than on the current length $L$ as in Eqs.~\ref{eq:hbar}, \ref{eq:gammabar}, and~\eqref{eq:phi-bar}. The corresponding dimensionless quantities are denoted by a superimposed tilde, namely
	\begin{equation}
		\tilde{h} = \frac{h}{L_{\rm max}} , \qquad
		\tilde{x} = \frac{x}{L_{\rm max}} , \qquad
		\tilde{L} = \frac{L}{L_{\rm max}} , \qquad
		\tilde{\gamma} = \frac{\gamma}{\alpha A L_{\rm max}} , \qquad
		\tilde{\Phi} = \frac{\Phi}{AL_{\rm max}} .
	\end{equation}

\begin{figure}
\centerline{
	\begin{tabular}{rrrr}
		{\footnotesize (a)}\!\!\! &
		& ~~{\footnotesize (b)}\!\!\! & \\[-2.5ex]
		& \inclps{!}{0.37\textwidth}{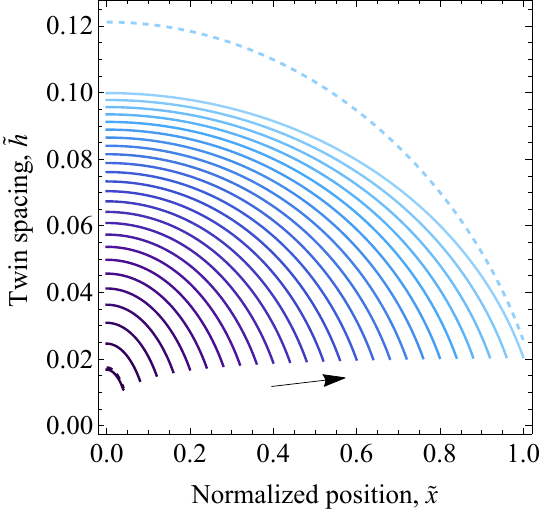} &
		& \inclps{!}{0.37\textwidth}{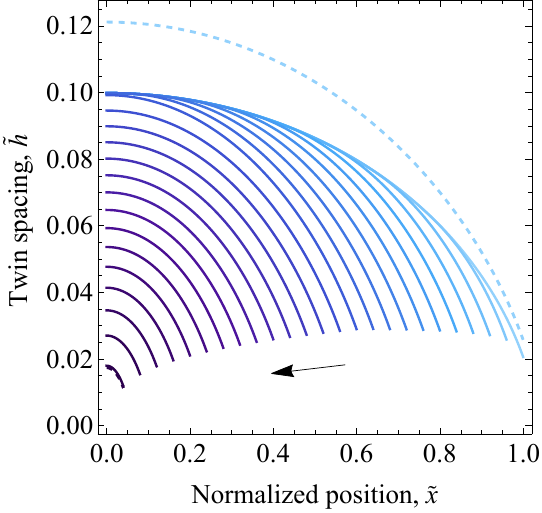} \\[2ex]
		{\footnotesize (c)}\!\!\! &
		& ~~{\footnotesize (d)}\!\!\! & \\[-2.5ex]
		& \inclps{!}{0.37\textwidth}{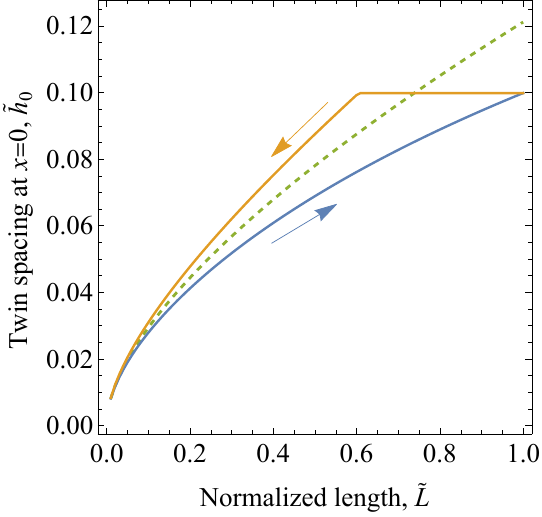} &
		& \inclps{!}{0.37\textwidth}{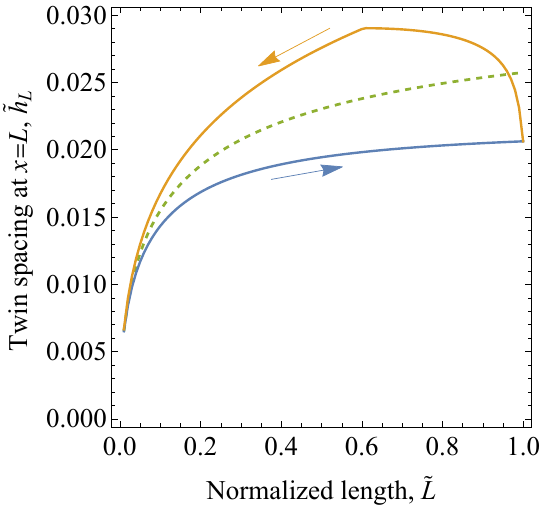}
	\end{tabular}
	}
\caption{Evolution of twin spacing $\tilde{h}$ for $\tilde{\gamma}=10^{-3}$ ($L_{\rm max}=20\,\mu$m) and $\bar{r}=0.21$ ($\hat{r}=4$\,MPa): (a,b) $\tilde{h}$ as a function of $\tilde{x}$ while $\tilde{L}$ is increasing (a) and decreasing (b); (c,d) twin spacing at $\tilde{x}=0$ (c) and at $\tilde{x}=\tilde{L}$ (d) as a function of the domain length $\tilde{L}$. The dashed lines correspond to the case with no dissipation, $\hat{r}=0$ {(in panels (a) and (b), the reference dashed lines are provided only for the smallest and largest $\tilde{L}$)}.}
\label{fig:diss:h:g10m3}
\end{figure}

Sample results are presented in Fig.~\ref{fig:diss:h:g10m3} for $L_{\rm max}=20\,\mu$m which corresponds to $\tilde{\gamma}=10^{-3}$ for the adopted material parameters. 
The evolution of twin spacing $\tilde{h}$ is shown in Fig.~\ref{fig:diss:h:g10m3}(a,b) where the individual curves correspond to selected values of the domain length $\tilde{L}$. As a reference, the dashed lines show the {profiles} of twin spacing obtained by minimization of the total free energy (no dissipation, $\hat{r}=0$) {for $\tilde{L}=0.04$ and for $\tilde{L}=1$ (the difference is significant only for $\tilde{L}=1$)}.

In the first part of the evolution, when the domain length $\tilde{L}$ increases (forward motion of the A--MM interface), the twin spacing increases, however, the increase is slowed down as compared to the case with no dissipation (dashed lines), see Fig.~\ref{fig:diss:h:g10m3}(a).

In the second part, when the domain length $\tilde{L}$ decreases (backward motion of the A--MM interface), Fig.~\ref{fig:diss:h:g10m3}(b), the evolution is more complex and comprises two stages. In the \emph{first stage} of the backward motion, the twinned domain is divided into two zones. In the outer zone, closer to the A--MM interface at $\tilde{x}=\tilde{L}$, the twin spacing evolves, while in the inner zone, closer to the free end at $\tilde{x}=0$, the twin spacing does not evolve, as the evolution is stopped by the dissipative forces. This can be seen in Fig.~\ref{fig:diss:h:g10m3}(b) where {the subsequent profiles overlap over a part of the domain}. 
The no-evolution zone gradually decreases during the first stage, and when it vanishes the \emph{second stage} of the backward motion starts, in which the twin spacing evolves in the entire domain.

The two-stage evolution of twin spacing during the backward motion can be seen in the plots of twin spacings $\tilde{h}_0$ and $\tilde{h}_L$, belonging to the ends, see Fig.~\ref{fig:diss:h:g10m3}(c,d). The transition from the first stage of the backward motion to the second one {is continuous but non-smooth, hence it} is marked by a kink in the evolution plots {(this is a consequence of adopting the non-smooth dissipation potential~\eqref{eq:Dri}, and the kink corresponds to an abrupt change of the rate of evolution).} In particular, at the free end, $\tilde{x}=0$, the twin spacing $\tilde{h}_0$ is constant during the first stage, i.e., until $\tilde{L}\approx0.6$ is reached, see Fig.~\ref{fig:diss:h:g10m3}(c). 
When $\tilde{L}$ decreases further, during the second stage, the twin spacing $\tilde{h}_0$ starts to decrease. 
At the same time, the twin spacing at the A--MM interface at $\tilde{x}=\tilde{L}$, increases during the first stage of the backward motion and decreases during the second stage.
Similar kinks are also seen in the plots of the evolution of the energy contributions that are discussed next.

\begin{figure}
\centerline{
	\begin{tabular}{rrrr}
		{\footnotesize (a)}\!\!\! &
		& ~~{\footnotesize (b)}\!\!\! & \\[-2.5ex]
		& \inclps{!}{0.37\textwidth}{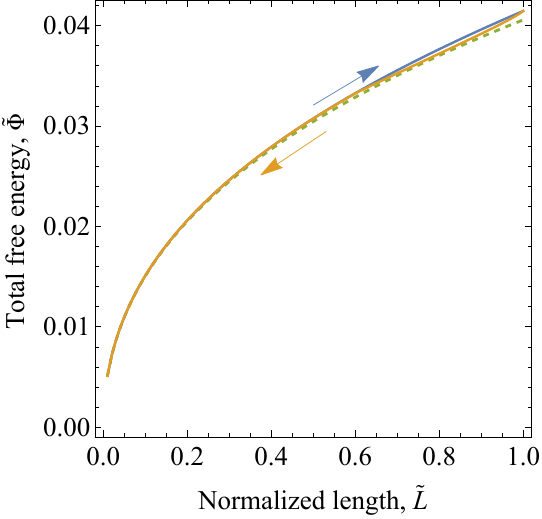} &
		& \inclps{!}{0.37\textwidth}{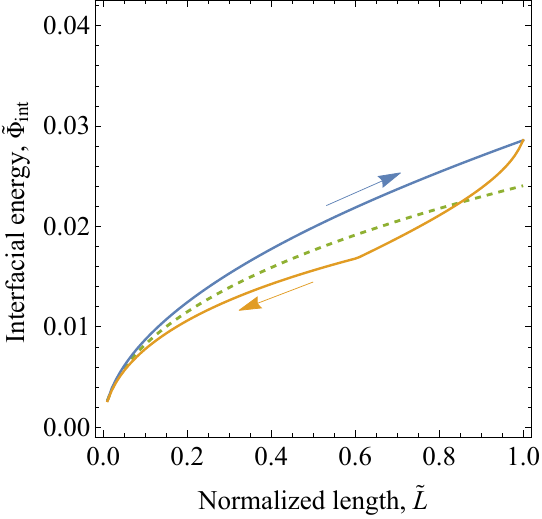} \\[2ex]
		{\footnotesize (c)}\!\!\! &
		& ~~{\footnotesize (d)}\!\!\! & \\[-2.5ex]
		& \inclps{!}{0.37\textwidth}{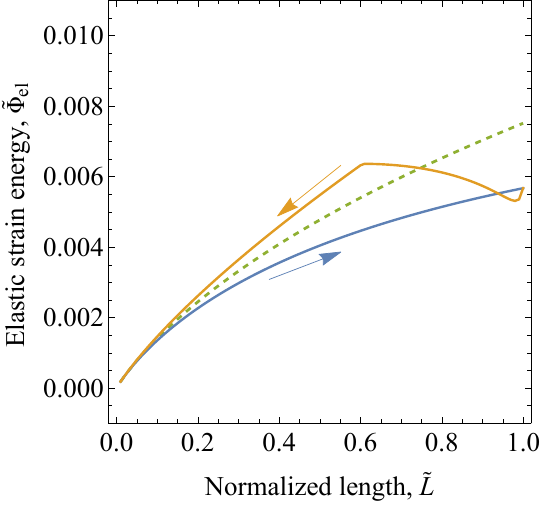} &
		& \inclps{!}{0.37\textwidth}{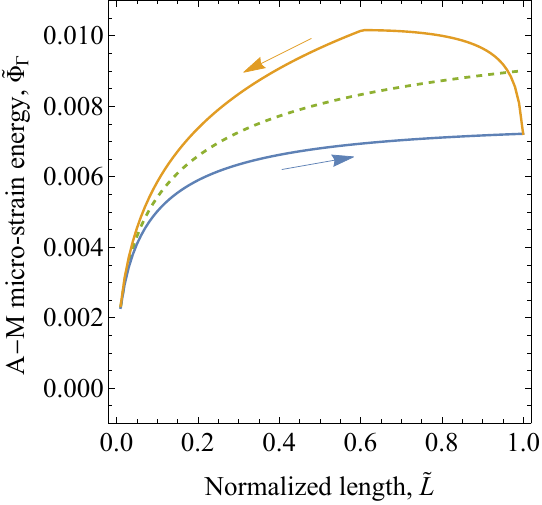}
	\end{tabular}
	}
\caption{Evolution of the total free energy $\tilde{\Phi}$ and individual energy contributions $\tilde{\Phi}_{\rm int}$, $\tilde{\Phi}_{\rm el}$, $\tilde{\Phi}_\Gamma$ for $\tilde{\gamma}=10^{-3}$ ($L_{\rm max}=20\,\mu$m) and $\bar{r}=0.21$ ($\hat{r}=4$\,MPa). The dashed lines correspond to the case with no dissipation, $\hat{r}=0$.}
\label{fig:diss:en:g10m3}
\end{figure}

Fig.~\ref{fig:diss:en:g10m3} shows the evolution of the total free energy $\tilde{\Phi}$ and its contributions defined in Eq.~\eqref{eq:Phi:contrib}. While the effect of energy dissipation on the individual contributions is significant, see Fig.~\ref{fig:diss:en:g10m3}(b,c,d), the total energy is only marginally affected and is very close to the case of no dissipation (depicted by the dashed line), see Fig.~\ref{fig:diss:en:g10m3}(a). 
Clearly, in the dissipative case, the total free energy is higher than in the case with no dissipation, since the later case corresponds to the minimum of the total free energy. 
As discussed above, the transition from the first stage of the backward motion of the A--MM interface to the second stage is associated with a kink at $\tilde{L}\approx0.6$ in each of the curves in Fig.~\ref{fig:diss:en:g10m3} (not visible in Fig.~\ref{fig:diss:en:g10m3}(a)).

\begin{figure}
\centerline{
	\begin{tabular}{rrrr}
		{\footnotesize (a)}\!\!\! &
		& ~~{\footnotesize (b)}\!\!\! & \\[-2.5ex]
		& \inclps{!}{0.37\textwidth}{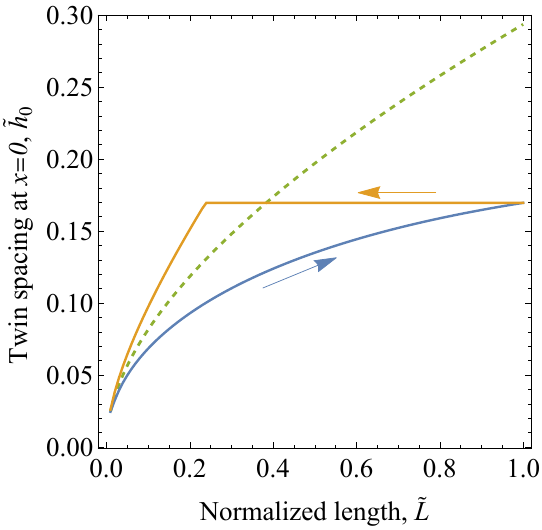} &
		& \inclps{!}{0.37\textwidth}{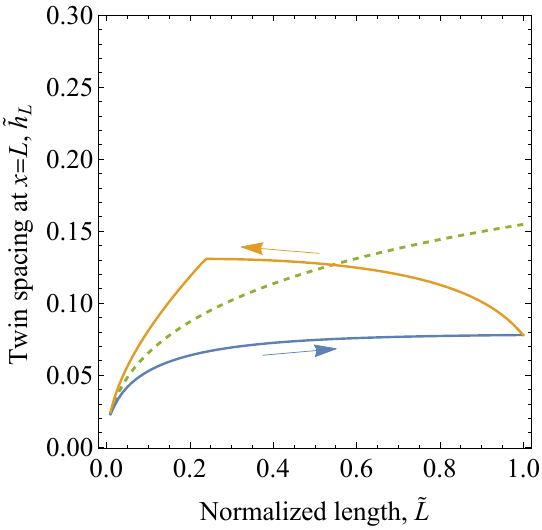} \\[2ex]
		{\footnotesize (c)}\!\!\! &
		& ~~{\footnotesize (d)}\!\!\! & \\[-2.5ex]
		& \inclps{!}{0.37\textwidth}{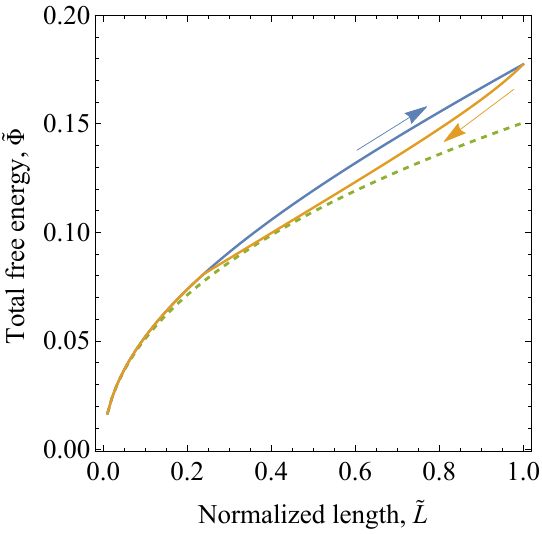} &
		& \inclps{!}{0.37\textwidth}{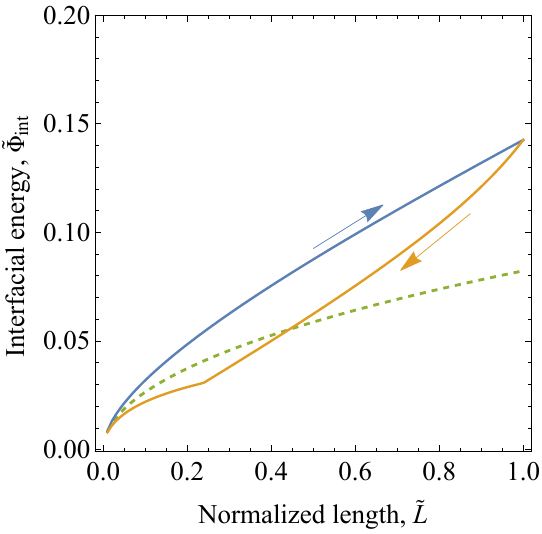} \\[2ex]
	\end{tabular}
	}
\caption{Evolution of the twin spacing $\tilde{h}$ at $\tilde{x}=0$ (a) and at $\tilde{x}=\tilde{L}$ (b), the total free energy $\tilde{\Phi}$ (c) and its interfacial contribution $\tilde{\Phi}_{\rm int}$ (d) for $\tilde{\gamma}=10^{-2}$ ($L_{\rm max}=2\,\mu$m) and $\bar{r}=0.21$ ($\hat{r}=4$\,MPa). The dashed lines correspond to the case with no dissipation, $\hat{r}=0$.}
\label{fig:diss:g10m2}
\end{figure}

\begin{figure}
\centerline{
	\begin{tabular}{rrrr}
		{\footnotesize (a)}\!\!\! &
		& ~~{\footnotesize (b)}\!\!\! & \\[-2.5ex]
		& \inclps{!}{0.37\textwidth}{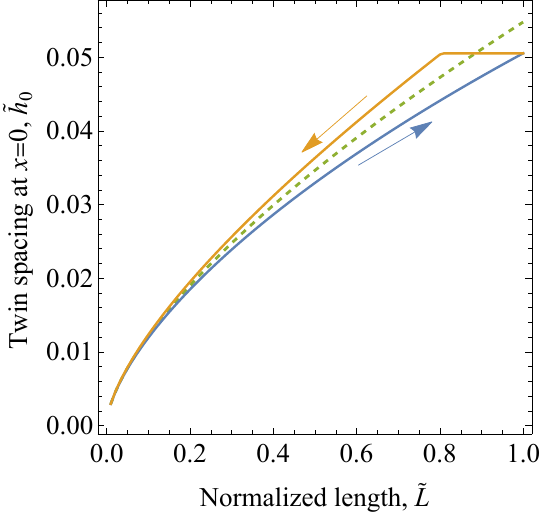} &
		& \inclps{!}{0.37\textwidth}{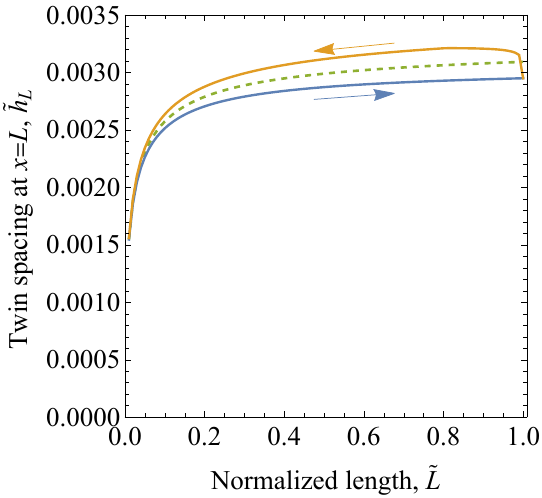} \\[2ex]
		{\footnotesize (c)}\!\!\! &
		& ~~{\footnotesize (d)}\!\!\! & \\[-2.5ex]
		& \inclps{!}{0.37\textwidth}{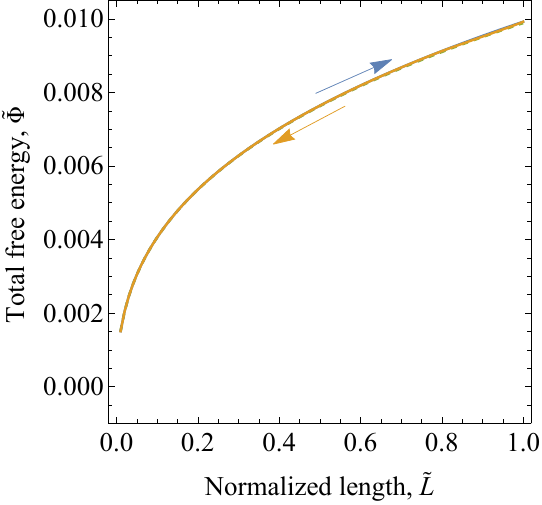} &
		& \inclps{!}{0.37\textwidth}{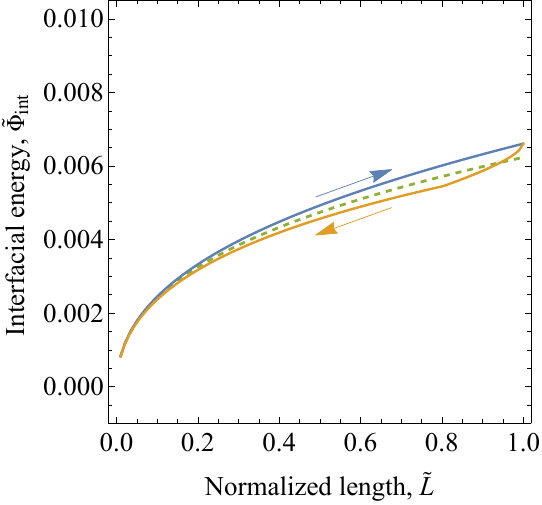} \\[2ex]
	\end{tabular}
	}
\caption{Evolution of the twin spacing $\tilde{h}$ at $\tilde{x}=0$ (a) and at $\tilde{x}=\tilde{L}$ (b), the total free energy $\tilde{\Phi}$ (c) and its interfacial contribution $\tilde{\Phi}_{\rm int}$ (d) for $\tilde{\gamma}=10^{-4}$ ($L_{\rm max}=200\,\mu$m) and $\bar{r}=0.21$ ($\hat{r}=4$\,MPa). The dashed lines correspond to the case with no dissipation, $\hat{r}=0$.}
\label{fig:diss:g10m4}
\end{figure}

To investigate the effect of energy dissipation further, two other cases are considered and reported in Figs.~\ref{fig:diss:g10m2} and~\ref{fig:diss:g10m4}. 
Specifically, the maximum domain length $L_{\rm max}$ is decreased and increased 10 times with respect to the previous example, while all the remaining parameters remain unchanged.
The case of $L_{\rm max}=2\,\mu$m, which corresponds to $\tilde{\gamma}=10^{-2}$, is shown in Fig.~\ref{fig:diss:g10m2}. It can be seen that the effect of dissipation on the evolution of twin spacing is in this case more pronounced than for $\tilde{\gamma}=10^{-3}$ shown in Fig.~\ref{fig:diss:h:g10m3}. Similarly, the effect on the total free energy $\tilde{\Phi}$ is also more pronounced, see Fig.~\ref{fig:diss:g10m2}(c). 
On the other hand, for $L_{\rm max}=200\,\mu$m, which corresponds to $\tilde{\gamma}=10^{-4}$, the effect of energy dissipation on the twin spacing is visible, but small, while the effect on the total free energy is negligible, see Fig.~\ref{fig:diss:g10m4}. 

The results presented in Figs.~\ref{fig:diss:h:g10m3}--\ref{fig:diss:g10m4}, correspond to a relatively large value of the rate-independent threshold ($\hat{r}=4$\,MPa) and to relatively small domain sizes ($L_{\rm max}$ varied between 2 and 200\,$\mu$m so that $\tilde{\gamma}$ varies between $10^{-2}$ and $10^{-4}$, respectively). For these parameters, the effect of energy dissipation on the evolution of branched microstructure, as characterized by the twin spacing, is visible, while the effect on the total free energy is limited (the effect is more pronounced for the individual energy contributions). 
For larger domain sizes $L_{\rm max}$ (i.e., for smaller values of the dimensionless twin boundary energy $\tilde{\gamma}$), the respective effects are smaller and can be considered insignificant. Of course, the effects of energy dissipation are also weaker for smaller values of the rate-independent threshold $\hat{r}$. For instance, when the case of $\tilde{\gamma}=10^{-2}$ is considered, as in Fig.~\ref{fig:diss:g10m2}, but parameter $\hat{r}$ is reduced twice (to 2\,MPa), the effect of energy dissipation is qualitatively similar (although somewhat more pronounced) to the case of $\tilde{\gamma}=10^{-3}$ and $\hat{r}=4$\,MPa, see Figs.~\ref{fig:diss:h:g10m3} and~\ref{fig:diss:en:g10m3}. 
The corresponding results are not shown.

\section{Discussion}
\label{sec:discussion}

The {model developed above is based on a simple estimate of the interfacial energy contribution, Eq.~\eqref{eq:Eint}, which leads to a particularly simple form of the corresponding term in the free energy density~\eqref{eq:phi}. This form corresponds to the assumption that the deviation of the orientation of the actual twin interfaces from the theoretical (nominal) orientation is small and thus negligible and, secondly, the local interfacial energy density is not affected by such small changes of the interface orientation. It is of interest to examine the validity of these assumptions.}

{Referring to Fig.~\ref{fig:speed}, the change of the interface orientation due to branching is characterized by angle $\beta$ which depends on the spatial derivative $h_{,x}$ of the twin spacing $h$, see Eq.~\eqref{eq:beta2}. The magnitude of the derivative $h_{,x}$ is the largest at $x=L$, see Fig.~\ref{fig:sample}, and is then equal to $\bar{\Gamma}=\Gamma/A$, according to the boundary condition~\eqref{eq:strong:bc}${}_2$. For the specific parameters adopted in Section~\ref{sec:predictions}, we have $\beta=5^\circ$ for $\bar{\Gamma}=1.2$ and $\beta=1.5^\circ$ for $\bar{\Gamma}=0.35$ so that the deviation of the orientation is indeed small. The corresponding change of the interface area is even smaller because the length of two twin interfaces (the upper ones in Fig.~\ref{fig:speed}(b)) increases, while the length of the other two interfaces decreases so that the net change is a second-order effect for small $\beta$.}

{Quantitative assessment of the effect of interface orientation on the interfacial energy is currently not possible since respective data is missing, and actually only rough estimates of the twin boundary energy itself are used. Here, atomistic simulations hold promise to deliver more reliable data, see e.g.\ recent results for NiTi \citep{larosa2024atomistic}. 
On the other hand, in a purely phenomenological manner, the related effects can be estimated using Eq.~\eqref{eq:Eint:full}. Adopting Eq.~\eqref{eq:Eint:full} instead of the simplified form in Eq.~\eqref{eq:Eint} and using Eqs.~\eqref{eq:Eel}${}_2$ and~\eqref{eq:dhdx}, the free energy density can be derived in the following extended form,
    \begin{equation}
        \label{eq:phi:ext}
        \phi(h,h_{,x}) = \frac12 A (h_{,x})^2 + \frac{2\gamma}{\alpha h} \left( 1 - \frac34 (1-\lambda) \lambda \alpha h_{,x} + \frac9{16} (1-\lambda)^2 (1+\lambda) \alpha^2 (h_{,x})^2 \right) ,
    \end{equation}
as a generalization of Eq.~\eqref{eq:phi}. The additional terms increase the interfacial energy contribution at $x=L$ by 14\% for $\bar{\Gamma}=1.2$ and by 4\% for $\bar{\Gamma}=0.35$, while the increase is smaller in the remaining part of the domain (the magnitude of $h_{,x}$ is the largest at $x=L$). However, the overall effect on the twin spacing is insignificant with the difference of less than 1\%, except in the vicinity of $x=L$, as our additional computations indicate (the finite-element implementation of the above extended model is straightforward and is not discussed here). Concluding, the effect of interface orientation on the interfacial energy can be disregarded.}

Following \citet{seiner2020branching}, the specific model discussed {in Sections~\ref{sec:model} and~\ref{sec:dissip}} corresponds to the situation in which the twinned domain is bounded by the A--MM interface on its one end and by a {parallel} free surface on the other end {(a non-parallel free surface can also be considered, as discussed below).}
This is a special case that may be encountered in a single crystal undergoing martensitic transformation. A different situation is expected in the case of a twinned martensite plate formed within a grain in a polycrystalline material. One possible scenario would be that the twinned martensite plate is bounded by two A--MM interfaces (no free surfaces). The proposed approach is directly applicable in such a case as well. It suffices to consider the domain $x\in[-L,L]$ with an adequate boundary condition at $x=-L$, namely $h_{,x}\big\vert_{x=-L}=\bar{\Gamma}/A$, cf.\ Eq.~\eqref{eq:strong:bc}${}_2$. Indeed, the corresponding solution can be obtained directly from the solution obtained for the original problem formulated in Section~\ref{sec:governing} by simply exploiting the symmetry with respect to $x=0$ (note the homogeneous natural boundary condition at $x=0$, Eq.~\eqref{eq:strong:bc}${}_1$).

An analogous extension of the problem in the case of evolution with dissipation effects would be possible only when the martensite plate grows and shrinks in a symmetric manner. Other situations, for instance, propagation of only one A--MM interface while the other one is halted, would require revision of the relationship between the rate of twin spacing and the local interface speed, see Eq.~\eqref{eq:vn}. 
Finally, several more complex situations may be encountered when twinned martensite domain terminates at other types of (macroscopic) interfaces, for instance, at the interface between two twinned domains or at the interface between twinned martensite and a single variant of martensite. Each case would require separate modeling assumptions.

{The assumption that the bounding surfaces are parallel, so that the length $L$ is independent of the position within the A--MM interface, can be relaxed, at least when the deviation is not too big. In such a case, the twin spacing $h$ would depend not only on the distance from the A--MM interface but also on the position within the A--MM interface, including in the third (out-of-plane) direction. Recall that the model assumes that there is no periodicity nor perfect arrangement of branching generations, hence individual twin lamellae branch independently and $h$ is an average twin spacing. The energetic cost of the additional in-plane variation of $h$ is thus expected to be insignificant. Accordingly, the 1D model can be solved locally at each position within the A--MM interface using the corresponding local length $L$. In practical terms, care would be needed here, since such a local 1D model would describe branching along an axis inclined with respect to the A--MM interface (parallel to the twin interfaces), while in the present formulation $L$ is measured in the direction normal to the A--MM interface. This is, however, only a technical detail.}

It has been shown in Section~\ref{sec:model} that the problem of minimization of the total free energy (no dissipation) is fully characterized by two dimensionless parameters, $\bar{\gamma}$ and $\bar{\Gamma}$, see Eqs.~\eqref{eq:strong:dimless} and~\eqref{eq:gammabar}. The results reported in Fig.~\ref{fig:compare} demonstrate that, over a wide range of values of the parameter $\bar{\gamma}$ (say, for $\bar{\gamma}<10^{-3}$), the dimensionless total energy $\bar{\Phi}$ and twin spacing $\bar{h}_0$ at $x=0$ are practically insensitive to $\bar{\Gamma}$ and hence depend solely on $\bar{\gamma}$. This regime corresponds to a significant activity of branching so that the twin spacing at the A--MM interface, $\bar{h}_L$, is much smaller than that at the free boundary, $\bar{h}_0$. Note, however, that the respective contribution to the total free energy, $\bar{\Phi}_\Gamma=\bar{\Gamma}\bar{h}_L$, is then small, but not negligible, see Fig.~\ref{fig:contributions}(b). Note also that, although $\bar{\Phi}$ and $\bar{h}_0$ are not sensitive to $\bar{\Gamma}$ in this regime, the twin spacing $\bar{h}_L$ is visibly affected by $\bar{\Gamma}$, and so is the number of branching generations, $N$, see Fig.~\ref{fig:compare}(c,d).

Given that the grain size in a polycrystalline aggregate typically ranges from 1\,$\mu$m to 100\,$\mu$m, the regime of small $\bar{\gamma}$ (equivalent to large $L$) is less relevant for microstructures developing in polycrystals. This is because $L$, which is bounded by the grain size, is then small (which corresponds to the regime of large $\bar{\gamma}$). Consequently, twin branching is less likely to occur in polycrystalline materials or is then less pronounced. However, this characterization is relative and also depends on specific material properties.

It is important to remark that there is quite some uncertainty concerning the values of the material parameters, particularly, the twin boundary energy $\gamma$ and the rate-independent interface propagation threshold $\hat{r}$. 
Also, the value of parameter $A$ characterizing the elastic strain energy of branching has been determined using the upper bound estimate provided by \citet{seiner2020branching}. To allow direct comparison with the predictions of the model of \citet{seiner2020branching}, we have used the estimate corresponding to isotropic elastic properties, Eq.~\eqref{eq:Eel}. On the other hand, it is well known that the CuAlNi alloy is elastically highly anisotropic \citep{sedlak2005elastic}, hence accounting for the elastic anisotropy may influence the value of $A$ (the corresponding estimate has also been provided by \citet{seiner2020branching}). In any case, the actual accuracy of those estimates is not known. Alternative estimates, for instance, from direct simulations of branched microstructures, are not available.

In the model of the evolution of the branched microstructure, we have assumed that dissipation is rate-independent. It is generally accepted that rate-independent dissipation is a relevant choice for shape memory alloys, at least, at higher spatial and temporal scales \citep{petryk2010interfacial:part1}. 
Concerning the rate-independent threshold $\hat{r}$, we have estimated a physically relevant range of its values using the experiments of \citet{novak2006transformation}. The availability of these estimates is actually another reason to adopt the rate-independent dissipation model. 
We have additionally developed a version of the model employing viscous dissipation, with the dissipation potential being quadratic in the interface speed $\hat{v}_{\rm n}$. We have verified that the viscous model can deliver results qualitatively similar to those of the rate-independent model. This is achieved by appropriately adjusting the interface mobility parameter (and/or the rate of change of the domain size $L$). Nevertheless, we do not report these results due to the lack of experimental evidence supporting a physically relevant value for interface mobility.

\section{Conclusion}
\label{sec:conclusion}

A new approach to the modeling of twin branching in shape memory alloys has been developed in which twin spacing is treated as a continuous function of the distance from the austenite--martensite interface. 
The free energy, which comprises the interfacial and elastic strain contributions, is then expressed in terms of the twin spacing and its gradient. The total free energy is minimized, and the problem is solved numerically using the finite element method. 
The resulting 1D model can be considered a continuous counterpart of the discrete model of \citet{seiner2020branching}, and a very good agreement between the two models has been demonstrated within the entire range of physically relevant model parameters.

The dimensionless formulation of the model, which in the continuous setting is introduced in a natural way, indicates that the problem is fully characterized by two dimensionless parameters expressed in terms of the material properties and of the size of the twinned domain.
The results show that, in a wide range of values of the dimensionless interfacial energy $\bar{\gamma}$, the solution is practically insensitive to parameter $\bar{\Gamma}$ characterizing the energy of elastic micro-strains at the A--MM interface. In this regime, the problem is thus fully characterized by only one dimensionless parameter, $\bar{\gamma}$.

The evolution of the branched microstructure has also been studied by considering the energy dissipation associated with propagation of the twin interfaces. 
The evolution is governed by a minimization problem formulated for a rate potential comprising the free energy rate and the dissipation potential. 
The results show that, for a high but realistic rate-independent threshold for interface propagation, the energy dissipation effects are significant only for high values of the dimensionless interfacial energy parameter $\bar{\gamma}$ (for instance, for a relatively small size of the twinned domain). Otherwise, the energy dissipation effects are negligible so that effectively the problem is history-independent and minimization of the free energy alone suffices to determine the twin spacing and other characteristics. 
Interestingly, the twin spacing and the individual contributions to the total free energy exhibit a visibly higher sensitivity to the dissipation effects than the total free energy itself.

To the best of our knowledge, this is the first time that the energy dissipation effects are included in the modeling of twin branching phenomena. This has been facilitated by the continuous description adopted in this work, as otherwise modeling of an evolution problem may not be readily feasible.

\paragraph{Acknowledgments} This work has been partially supported by the National Science Centre (NCN) in Poland through the Grant No.\ 2021/43/D/ST8/02555.

\appendix

\setcounter{figure}{0}

\section{Analytical solution for the static case}
\label{app:analytical}

For completeness, an analytical solution for the static case presented in Section~\ref{sec:model} is provided in this appendix. The derivation follows that reported by \citet{zhang2024austenite}. The difference is that we exploit the boundary condition at $x=L$, Eq.~\eqref{eq:strong:bc}${}_2$, which is not considered by \citet{zhang2024austenite}. Secondly, we provide a full final form without a simplifying assumption, as adopted in their work. This is commented at the end of the appendix.

The derivations are carried out in a dimensionless form. The starting point is the equation of equipartition of energy~\eqref{eq:equipart},
	\begin{equation}
		\label{eq:equipart:dl}
		\frac12 (\bar{h}_{,\bar{x}})^2 - \frac{2\bar{\gamma}}{\bar{h}} = \bar{C} .
	\end{equation}
We note that at $\bar{x}=0$ we have $\bar{h}_{,\bar{x}}=0$, cf.\ Eq.~\eqref{eq:strong:dimless}${}_1$, which allows us to determine the constant $\bar{C}=-2\bar{\gamma}/\bar{h}_0$ so that 
Eq.~\eqref{eq:equipart:dl} can be rewritten in the following form
	\begin{equation}
            \label{eq:app:dhdx}
		\frac{\rd \bar{h}}{\rd \bar{x}} = -2 \sqrt{ \bar{\gamma} \left( \frac{1}{\bar{h}} - \frac{1}{\bar{h}_0} \right) } .
	\end{equation}
Here, we anticipate that the twin spacing is a monotonically decreasing function, hence the minus sign on the right-hand side. 

Let us now consider $\bar{x}$ to be a function of $\bar{h}$ so that Eq.~\eqref{eq:app:dhdx} implies
	\begin{equation}
		\frac{\rd \bar{x}}{\rd \bar{h}} = -\frac12 \sqrt{ \frac{\bar{h}_0 \bar{h}}{\bar{\gamma}(\bar{h}_0-\bar{h})} } .
	\end{equation}
This equation can be integrated analytically, thus yielding
	\begin{equation}
		\bar{x}(\bar{h}) = \frac12 \sqrt{ \frac{\bar{h}_0^3}{\bar{\gamma}} } \left( \sqrt{ \frac{\bar{h}}{\bar{h}_0} \left( 1 - \frac{\bar{h}}{\bar{h}_0} \right) } - \arcsin \sqrt{\frac{\bar{h}}{\bar{h}_0}} \right) + \bar{B} .
	\end{equation}
The integration constant $\bar{B}$ can be determined from the condition that $\bar{x}(\bar{h}_0)=0$, which gives
	\begin{equation}
		\bar{B} = \frac{\pi}{4} \sqrt{ \frac{\bar{h}_0^3}{\bar{\gamma}} } .
	\end{equation}
Finally, we have
	\begin{equation}
		\bar{x}(\bar{h}) = \frac12 \sqrt{ \frac{\bar{h}_0^3}{\bar{\gamma}} } \left( \sqrt{ \frac{\bar{h}}{\bar{h}_0} \left( 1 - \frac{\bar{h}}{\bar{h}_0} \right) } + \frac{\pi}{2} - \arcsin \sqrt{\frac{\bar{h}}{\bar{h}_0}} \right) .
	\end{equation}
In this formula, $\bar{h}_0$ is unknown and can be determined from the boundary conditions at $\bar{h}=\bar{h}_L$, i.e., at the A--MM interface. Since $\bar{h}_L$ is also unknown, we have two unknowns ($\bar{h}_0$, $\bar{h}_L$) and two equations,
	\begin{equation}
            \label{eq:xh}
		\bar{x}(\bar{h}_L) = 1 , \qquad 
            \frac{\rd \bar{x}}{\rd \bar{h}} (\bar{h}_L) = -\frac{1}{\bar{\Gamma}}.
	\end{equation}
These equations must be solved numerically.

Note that \citet{zhang2024austenite} adopt a simplifying assumption concerning the integration constant $\bar{B}$, namely $\bar{B}\approx1$, so that their Eq.~(11) takes a particularly simple form. While this assumption may be a reasonable approximation in some cases, it does not hold in general, i.e., in the entire range of physically relevant values of the governing parameters $\bar{\gamma}$ and $\bar{\Gamma}$.

\section{Estimation of local interface propagation speed}
\label{app:speed}

\begin{figure}
	\centerline{\inclps{0.8\textwidth}{!}{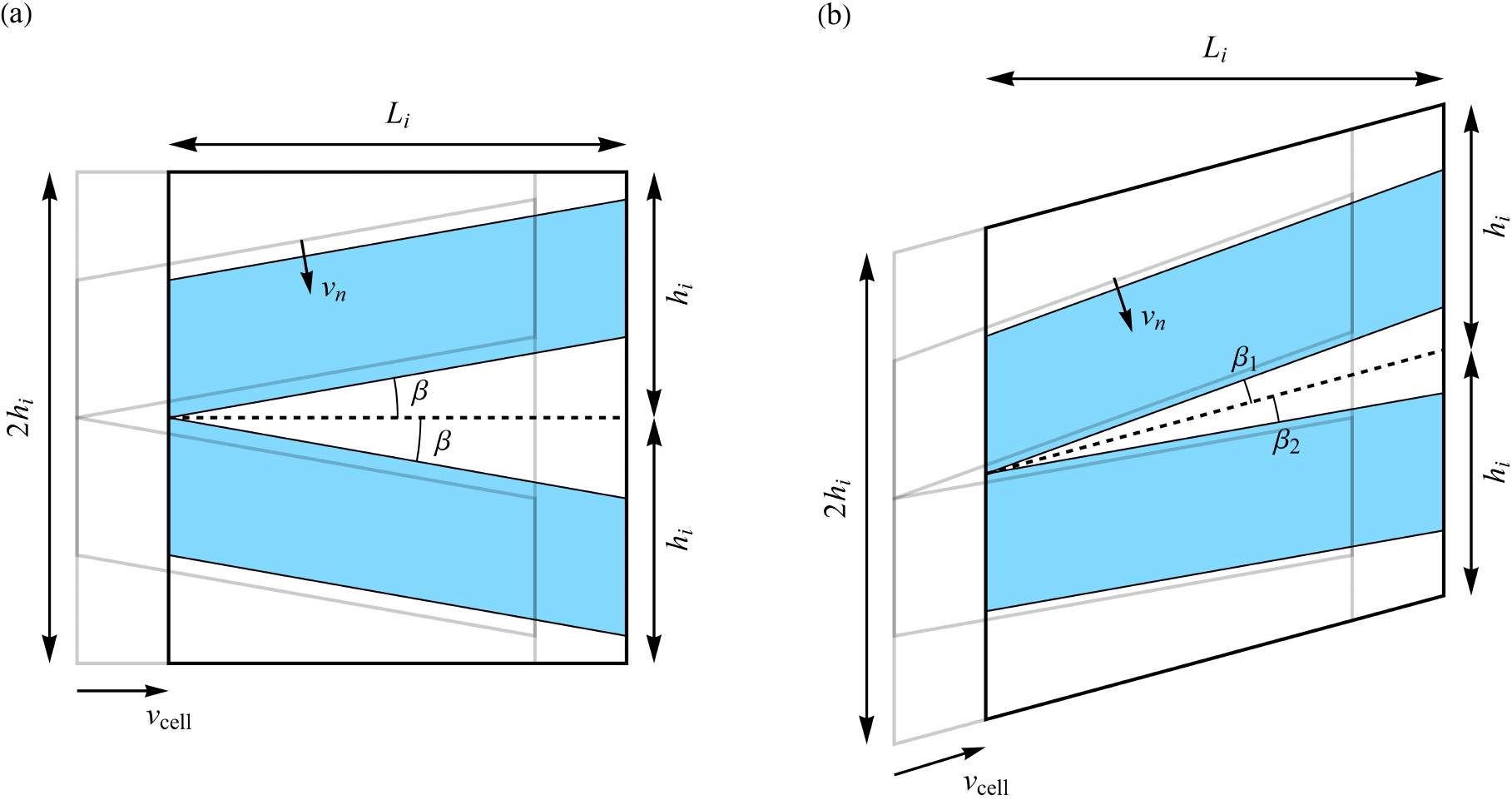}}
	\caption{Estimating the local interface propagation speed $\hat{v}_{\rm n}$: (a) $\alpha=1$; (b) $\alpha\neq1$. The location of the cell at an earlier instant is indicated by gray lines.}
	\label{fig:speed}
\end{figure}

In this appendix, an estimate of the local interface propagation speed $\hat{v}_{\rm n}$ is derived as a function of $\dot{h}
$, the rate of change of the twin spacing. 
Referring to the discrete model of branching shown in Fig.~\ref{fig:Seiner}, we assume that locally the evolution of the branched microstructure, i.e., the change of the twin spacing, proceeds by a `rigid' translation of the cells, see Fig.~\ref{fig:speed}. In this derivation, we neglect the possible change of the proportions of the cells.

The derivation is first performed for a simplified case where the (nominal) twin interfaces are perpendicular to the A--MM interface, thus $\bfm{m}\cdot\bfm{n}=0$ and $\alpha=1$, see Eq.~\eqref{eq:alpha}${}_1$. 
Consider thus a rigid translation of the unit cell with the speed $v_{\rm cell}$, as shown in Fig.~\ref{fig:speed}(a). The resulting interface propagation speed $\hat{v}_{\rm n}$ (measured in the direction normal to the local twin interfaces) is thus given by
	\begin{equation}
		\label{eq:vcell1}
		\hat{v}_{\rm n} = v_{\rm cell} \sin\beta \approx v_{\rm cell} \tan\beta.
	\end{equation}
Here and below, we assume that $L_i\gg h_i$ which implies that the angle $\beta$ is small. From the geometry of the unit cell, we have
	\begin{equation}
		\label{eq:beta1}
		\tan\beta = \frac{(1-\lambda) h_i}{2} \frac{1}{L_i} = - \frac34 (1-\lambda) h_{,x} \qquad \mbox{(for $\alpha=1$)} ,
	\end{equation}
where Eq.~\eqref{eq:dhdx} has also been used.

In line with the assumption of rigid translation of the unit cell, the `material time derivative' of the twin spacing $h=h(x,t)$ vanishes so that
	\begin{equation}
		\label{eq:DhDt1}
		\frac{{\rm D} h}{{\rm D} t} = 
			\frac{\partial h}{\partial t} + \frac{\partial h}{\partial x} \frac{\partial x}{\partial t} =
			\dot{h} + h_{,x} v_{\rm cell} = 0 , \qquad
		v_{\rm cell} = - \frac{\dot{h}}{h_{,x}} \qquad \mbox{(for $\alpha=1$)} .
	\end{equation}
Combining Eqs.~\eqref{eq:vcell1}--\eqref{eq:DhDt1}, the interface speed $\hat{v}_{\rm n}$ can now be expressed as a function of $\dot{h}$,
	\begin{equation}
		\label{eq:vn1}
		\hat{v}_{\rm n} = \frac34 (1-\lambda) \dot{h} \qquad \mbox{(for $\alpha=1$)} .
	\end{equation}

Let us now consider the general case of $\bfm{m}\cdot\bfm{n}\neq0$ and $\alpha\neq1$, see Fig.~\ref{fig:speed}(b). The unit cell is no longer symmetric. This in particular implies that angles $\beta_1$ and $\beta_2$ are not equal. However, in view of $L_i\gg h_i$, we neglect this (small) difference, and the counterpart of Eq.~\eqref{eq:beta1} reads 
	\begin{equation}
		\label{eq:beta2}
		\beta_1 \approx \beta_2 = \beta , \qquad
		\tan\beta = \frac{(1-\lambda) \alpha h_i}{2} \frac{\alpha}{L_i} = - \frac34 (1-\lambda) \alpha^2 h_{,x} .
	\end{equation}
This also implies that the propagation speed $\hat{v}_{\rm n}$ is equal for all interfaces and that Eq.~\eqref{eq:vcell1} holds.

Considering that the rigid translation of the unit cell is directed along the nominal twin interfaces, the `advection' speed in Eq.~\eqref{eq:DhDt1} is now equal to $\alpha v_{\rm cell}$ so that the counterpart of Eq.~\eqref{eq:DhDt1} takes the form
	\begin{equation}
		\label{eq:DhDt2}
		\frac{{\rm D} h}{{\rm D} t} = 
			\dot{h} + h_{,x} \alpha v_{\rm cell} = 0 , \qquad
		v_{\rm cell} = - \frac{\dot{h}}{\alpha h_{,x}} .
	\end{equation}

By combing Eqs.~\eqref{eq:vcell1}, \eqref{eq:beta2} and~\eqref{eq:DhDt2} the dependence of the interface speed $\hat{v}_{\rm n}$ on the rate of change of the twin spacing, $\dot{h}$, is finally found in the following form,
	\begin{equation}
		\label{eq:vn2}
		\hat{v}_{\rm n} = \frac34 (1-\lambda) \alpha \dot{h} .
	\end{equation}

Refined estimates could be derived by considering the actual geometry in more detail. The simple estimate derived above is sufficient for the present purpose, as it shows the structure of the dependence of $\hat{v}_{\rm n}$ on $\dot{h}$.

\section{Computational treatment of the evolution problem}
\label{app:AL}

The minimization problem~\eqref{eq:min} is non-smooth, and its computational treatment relies on the augmented Lagrangian technique~\citep{alart1991mixed}, borrowed here from computational contact mechanics. 
An equivalent smooth saddle-point problem is thus formulated,%
	\begin{equation}
		\label{eq:minmax}
		{\cal L} = \dot{\Phi} + \Lambda \;\to\; \min_{\dot{h}} \max_{\vphantom{\dot{h}}\lambda} ,
	\end{equation}
where $\lambda$ denotes (in this appendix only) the Lagrange multiplier field that is introduced through the functional $\Lambda$,
	\begin{equation}
		\label{eq:cL}
		\Lambda = \int_0^L \frac{r}{h} \, l(\dot{h},\lambda) \rd x .
	\end{equation}
The function $l(\dot{h},\lambda)$ is defined as
	\begin{equation}
		\label{eq:lAL}
		l(\dot{h},\lambda) = \left\{
			\begin{array}{ll}
				\left( \lambda+ \frac12 \varrho \dot{h} \right) \dot{h} & \mbox{if} \;\; |\hat{\lambda}| \leq 1 , \\[1ex]
				-\frac{1}{2\varrho} \left(\lambda^2 - 2 | \hat{\lambda} | +1 \right) \; & \mbox{if} \;\; |\hat{\lambda}| > 1 ,
			\end{array}
			\right.
	\end{equation}
where $\hat{\lambda}=\lambda+\varrho \dot{h}$ is the augmented Lagrange multiplier, and $\varrho>0$ is a positive constant.

The Lagrange functional ${\cal L}$ is smooth (because function $l(\dot{h},\lambda)$ is continuously differentiable) hence the solution of the saddle-point problem~\eqref{eq:minmax} can be found by solving the condition of stationarity of ${\cal L}$ with respect to arbitrary variations $\delta \dot{h}$ and $\delta \lambda$, which reads
	\begin{eqnarray}
		\int_0^L \left( A h_{,x} \delta\dot{h}_{,x} - \frac{2\gamma}{\alpha h^2} \delta\dot{h} +
		\frac{r}{h} \hat{\lambda}_{\rm eff} \, \delta\dot{h} \right) \rd x 
		+ \Gamma \, \delta\dot{h} \big\vert_{x=L}
		\,=\, 0 & &
		\quad \forall \, \delta \dot{h} ,
		\label{eq:weak:AL:h}
		\\
		\label{eq:weak:AL:lambda}
		\int_0^L \frac{r}{h} C_{\rm eff} \, \delta \lambda \, \rd x 
		\,=\, 0 & &
		\quad \forall \, \delta \lambda ,
	\end{eqnarray}
where
	\begin{equation}
		\label{eq:dlAL}
		\hat{\lambda}_{\rm eff} = \frac{\partial l}{\partial \dot{h}} = \left\{
			\begin{array}{ll}
				\hat{\lambda} & \mbox{if} \;\; |\hat{\lambda}| \leq 1 , \\[1ex]
				\sign \hat{\lambda} \; & \mbox{if} \;\; |\hat{\lambda}| > 1 ,
			\end{array}
			\right.
		\qquad
		C_{\rm eff} = \frac{\partial l}{\partial \lambda} = \left\{
			\begin{array}{ll}
				\dot{h} & \mbox{if} \;\; |\hat{\lambda}| \leq 1 , \\[1ex]
				-\frac{1}{\varrho} \left( \lambda - \sign \hat{\lambda} \right) \; & \mbox{if} \;\; |\hat{\lambda}| > 1 .
			\end{array}
			\right.
	\end{equation}
Eq.~\eqref{eq:weak:AL:h} is a smooth counterpart of Eq.~\eqref{eq:weak:ri} with the effective Lagrange multiplier $\hat{\lambda}_{\rm eff}$ replacing the subdifferential $\partial_{\dot{h}}|\dot{h}|$. On the other hand, Eq.~\eqref{eq:weak:AL:lambda} is a weak form of a state-dependent constraint $C_{\rm eff}=0$ which enforces $\dot{h}=0$ when $|\hat{\lambda}| \leq 1$ and $\lambda=\sign \hat{\lambda}=\pm1$ when $|\hat{\lambda}| > 1$. 
Note that the augmented Lagrangian method delivers a numerically exact solution that is independent of the value of parameter $\varrho$. 

At each time increment, the saddle-point problem specified above is solved using the finite element method. Recall that to facilitate enforcement of the natural boundary condition at $x=L$, see Eq.~\eqref{eq:strong:bc}${}_2$, the computations are carried out in a dimensionless form on a fixed domain $0\leq\bar{x}\leq1$ (and on a fixed mesh). 
Applying the backward-Euler time integration scheme, the dimensionless form of Eqs.~\eqref{eq:weak:AL:h} and~\eqref{eq:weak:AL:lambda} reads
	\begin{eqnarray}
		\int_0^1 \left( \bar{h}_{,\bar{x}} \delta\bar{h}_{,\bar{x}} - \frac{2\bar{\gamma}}{\bar{h}^2} \delta\bar{h} +
		\frac{\bar{r}}{\bar{h}} \bar{\lambda}_{\rm eff} \, \delta\bar{h} \right) \rd \bar{x} 
		+ \bar{\Gamma} \, \delta\bar{h} \big\vert_{\bar{x}=1}
		\,=\, 0 & &
		\quad \forall \, \delta \bar{h} ,
		\label{eq:weak:AL:h:dl}
		\\
		\label{eq:weak:AL:lambda:dl}
		\int_0^1 \frac{\bar{r}}{\bar{h}} \bar{C}_{\rm eff} \, \delta \lambda \, \rd \bar{x} 
		\,=\, 0 & &
		\quad \forall \, \delta \lambda ,
	\end{eqnarray}
where
	\begin{equation}
		\label{eq:dlAL:dl}
		\bar{\lambda}_{\rm eff} = \left\{
			\begin{array}{ll}
				\bar{\lambda} & \mbox{if} \;\; |\bar{\lambda}| \leq 1 , \\[1ex]
				\sign \bar{\lambda} \; & \mbox{if} \;\; |\bar{\lambda}| > 1 ,
			\end{array}
			\right.
		\qquad
		\bar{C}_{\rm eff} = \left\{
			\begin{array}{ll}
				\Delta\bar{h} & \mbox{if} \;\; |\bar{\lambda}| \leq 1 , \\[1ex]
				-\frac{1}{\bar{\varrho}} \left( \lambda - \sign \bar{\lambda} \right) \; & \mbox{if} \;\; |\bar{\lambda}| > 1 ,
			\end{array}
			\right.
	\end{equation}
$\bar{\lambda}=\lambda+\bar{\varrho}\Delta\bar{h}$ and $\bar{\varrho}=L\varrho/\tau$. 
Note that the problem is rate-independent, hence, upon time discretization, the rate $\dot{h}$ is here approximated directly by the increment $\Delta h=h_{n+1}-h_n$ without specifying the corresponding time increment, and a constant $\tau=1$\,s has been introduced to preserve the consistency of units.

As already discussed in Section~\ref{sec:evolution}, in the dimensionless setting, the rate (or increment) of the twin spacing must be evaluated at a fixed coordinate $x$. 
Accordingly, at each point $\bar{x}=x/L_{n+1}$, the increment $\Delta\bar{h}$ is defined as
	\begin{equation}
		\Delta\bar{h}(\bar{x}) = \frac{h_{n+1}(x)-h_n^{\rm ext}(x)}{L_{n+1}} = \bar{h}_{n+1}(\bar{x}) - \frac{L_n}{L_{n+1}} \bar{h}_n^{\rm ext}(\bar{x}_n^\ast)
		, \qquad
		\bar{x}_n^\ast = \frac{L_{n+1}}{L_n} \bar{x} ,
	\end{equation}
where $\bar{x}_n^\ast$ is the dimensionless position of the point $x=\bar{x}L_{n+1}$ at the previous time step and
	\begin{equation}
		\bar{h}_n^{\rm ext} (\bar{x}) = \left\{
			\begin{array}{ll}
				\bar{h}_n (\bar{x}) \;\; & \mbox{if} \;\; 0\leq\bar{x}\leq1 , \\
				0 & \mbox{if} \;\; \bar{x}>1 .
			\end{array}
			\right.
	\end{equation}
Here, $\bar{h}_n^{\rm ext}$ is an extension of $\bar{h}_n$ defined also for $\bar{x}>1$, which is needed to compute $\Delta\bar{h}$ when $L_{n+1}>L_n$. 
The above simple treatment is sufficient because, in the case of rate-independent dissipation, only the sign of the increment $\Delta\bar{h}$ enters the actual governing equations.

The weak forms~\eqref{eq:weak:AL:h:dl} and~\eqref{eq:weak:AL:lambda:dl} constitute the basis for the finite-element implementation of the model. 
The Lagrange multipliers $\lambda$ are introduced as additional global unknowns, otherwise the implementation follows a standard procedure, as briefly described in Section~\ref{sec:FE}.

\bibliographystyle{elsarticle-harv}
\bibliography{bibliography}

\end{document}